\documentclass[bibyear]{aa}  

\usepackage{graphicx}
\usepackage[dvipsnames]{xcolor}
\usepackage{marvosym}
\usepackage{natbib}
\bibpunct{(}{)}{;}{a}{}{,} 
\usepackage{xspace}
\usepackage[colorlinks=true, linkcolor=blue, citecolor=blue, urlcolor=blue]{hyperref}

\usepackage{subcaption,booktabs}
\usepackage{comment}
\usepackage[thinc]{esdiff}
\newcommand{\msun}{{\rm M}_\odot}

\newcommand{\petar}{\textsc{petar}\xspace}
\newcommand{\mobse}{\textsc{mobse}\xspace}
\newcommand{\galpy}{\textsc{galpy}\xspace}

\usepackage{soul}
\usepackage{txfonts}
\usepackage{pifont}

\newcommand{\orcidicon}[1]{\href{https://orcid.org/#1}{\includegraphics[width=11pt]{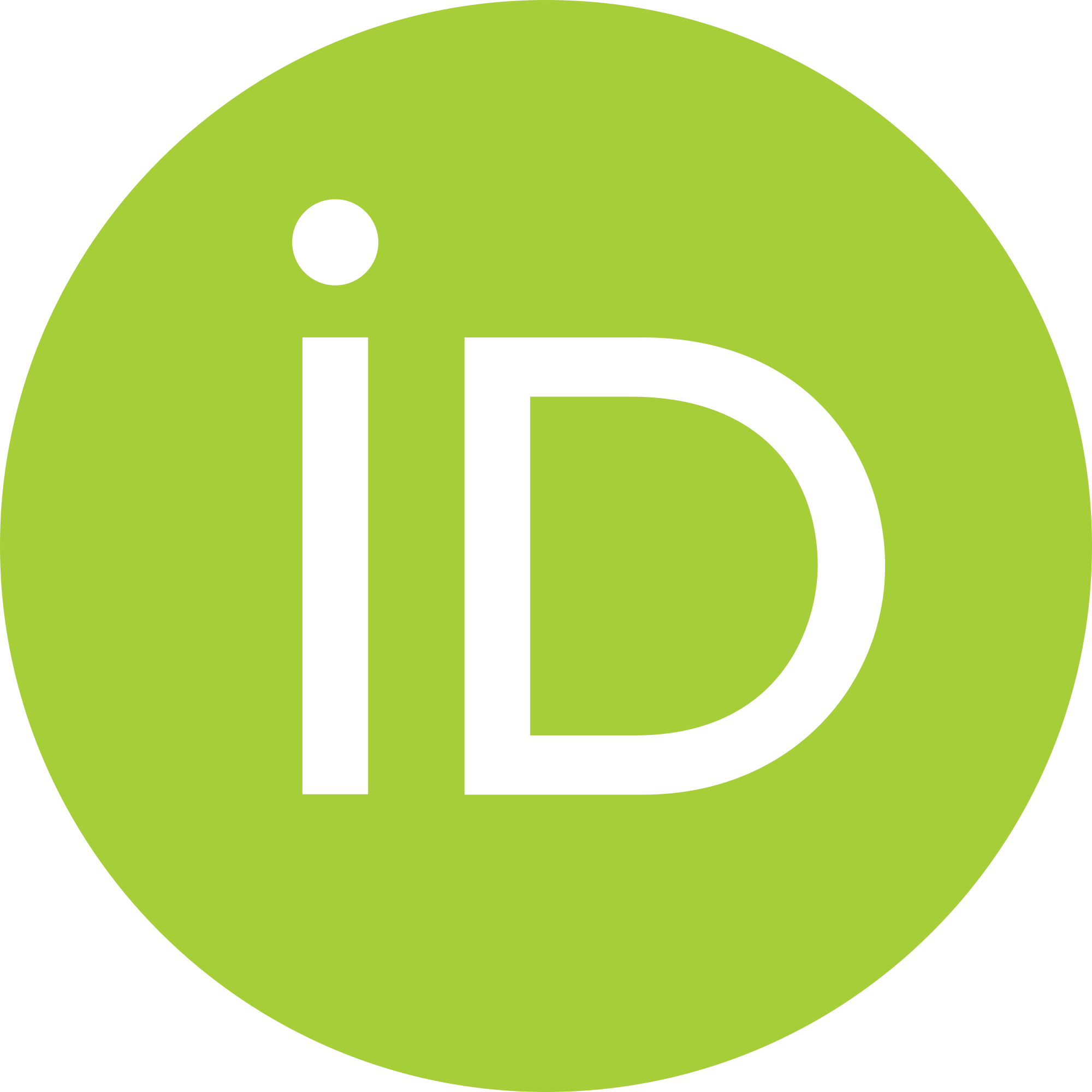}}}
\newcommand{\orcid}[1]{\href{https://orcid.org/#1}{\protect\orcidicon{#1}}}
\begin{document}

   \title{\emph{Teen} \textsc{TITANS} simulations - I. Inefficient intermediate-mass black hole seeding via stellar collisions in young massive clusters}

   \authorrunning{B. Mestichelli et al.}
   \titlerunning{\emph{Teen} TITANS simulations - I}
   \author{Benedetta Mestichelli
          \inst{1,2,3}
          \orcid{0009-0002-1705-4729} 
          \thanks{\href{mailto:benedetta.mestichelli@gssi.it}{benedetta.mestichelli@gssi.it}}
          \and 
          Sara Rastello\inst{4,5}\orcid{0000-0002-5699-5516}
          \thanks{\href{mailto:sara.rastello@fqa.ub.edu}{sara.rastello@fqa.ub.edu}}
          \and
          Michela Mapelli\inst{3,6,7,8}\orcid{0000-0001-8799-2548} \thanks{\href{mailto:mapelli@uni-heidelberg.de}{mapelli@uni-heidelberg.de}}
          \and 
          Manuel Arca Sedda\inst{1,2,9}\orcid{0000-0002-3987-0519}
          \and \\
          Marica Branchesi\inst{1,9}
          }
          
          \authorrunning{B. Mestichelli et al.}
          \institute{
          Gran Sasso Science Institute (GSSI), Viale Francesco Crispi 7, 67100, L’Aquila, Italy
          \and
          INFN, Laboratori Nazionali del Gran Sasso, I-67100 Assergi, Italy
           \and
          Institut f\"ur Theoretische Astrophysik, Zentrum f\"ur Astronomie, Universit\"at Heidelberg, Albert Ueberle Str. 2, D-69120 Heidelberg, Germany
           \and
        Departament de Física Quàntica i Astrofísica (FQA), Universitat de Barcelona (UB), c. Martí i Franquès 1, 08028 Barcelona, Spain
        \and
        Institut de Ciències del Cosmos (ICCUB), Universitat de Barcelona (UB), c. Martí i Franquès 1, 08028 Barcelona, Spain
          \and
          Universit\"at Heidelberg, Interdisziplin\"ares Zentrum f\"ur Wissenschaftliches Rechnen, Heidelberg, Germany
          \and
          Dipartimento di Fisica e Astronomia Galileo Galilei, Università di Padova, Vicolo dell’Osservatorio 3, I–35122 Padova, Italy
           \and
           INFN-Padova, Via Marzolo 8, I–35131 Padova, Italy
        \and
        INAF Osservatorio Astronomico d'Abruzzo, Via Maggini, 64100 Teramo, Italy  
          }
    
   \date{}

\abstract{Young massive clusters (YMCs) are dense stellar systems that provide favorable
environments for frequent stellar collisions, potentially leading to the formation of very massive stars (VMSs) and seeds of intermediate-mass black holes (IMBHs). In this work, we investigate the role of repeated stellar collisions in YMCs using \textsc{titans}, a new suite of 18 $N$-body simulations. Our models span cluster masses of $10^{5}-10^{6}\,\msun$,
half-mass densities $\rho_{\rm h}=100-10^{5}\,\msun\,\rm pc^{-3}$, and include high primordial binary fractions, consistent with observations of massive stars in young clusters. 
Overall, our simulations assume cluster properties that are typical of YMCs in the low-redshift Universe. We find that repeated stellar collisions are efficient only in the densest clusters with short relaxation times and do not occur in systems with $\rho_{\rm h}\lesssim500\,\msun\,\rm pc^{-3}$ and half-mass relaxation times $t_{\rm rh}\gtrsim1.3\,\rm Gyr$. Rapid mass segregation allows the most massive stars to sink toward the cluster center, merge, and undergo subsequent collisions, even in clusters with long core-collapse times. Nevertheless, stellar collision chains are typically triggered by the merger of a primordial binary and most often involve only two collisions. In our simulations, only three VMSs form via repeated stellar collisions and reach masses $m_*>330\,\msun$, while the majority has $m_*<300\,\msun$ and forms via primordial binary mergers. None of these objects represent viable IMBH seeds, as most of them develop helium cores in the (pulsational) pair-instability regime. We form five IMBHs from single or repeated stellar collisions involving stars at different evolutionary stages, while the dominant IMBH formation channel remains the merger of stellar-mass black holes, which produces twelve IMBHs. For densities and half-mass radii typical of local YMCs, our models show that stellar collision chains are inefficient in producing IMBHs more massive than $140\,\msun$, as most collisionally formed VMSs attain masses that fall in the pair-instability regime.}

\maketitle

\section{Introduction}

Young massive clusters (YMCs) are dense stellar systems with typical masses $M_{\rm cl}\gtrsim10^{4}\,\msun$, half-mass radii of a few parsecs, and ages $\lesssim100\,\rm Myr$ \citep{pzwart2010, bastian2013}. They are observed in a wide range of environments, from nearby star-forming galaxies to extreme starbursts, and are thought to represent the progenitors of globular clusters \citep{krumholz2019}. Owing to their high stellar densities and large populations of massive stars, YMCs provide key laboratories for studying stellar dynamics, binary evolution, and the formation of compact objects, including stellar-mass black holes (BHs), gravitational-wave (GW) sources, and possible seeds of intermediate-mass black holes (IMBHs) with $m_{\rm BH}\sim10^{2}-10^{5}\,\msun$ \citep{pzwart2002, pzwart2004, mapelli2016, dicarlo2019, dicarlo2020, kremer2020b, dicarlo2021, rastello2020, rastello2021, mas2021,
rantala2024, rantala2025, vergara2025}.

In sufficiently dense clusters, stellar collisions are expected to occur frequently, particularly among massive stars that rapidly segregate toward the cluster center. Early studies showed that such collisions may trigger runaway growth, leading to the formation of very massive stars (VMSs) and, potentially,
IMBHs. This process is especially relevant in clusters with short core-collapse times, where collisions are expected to begin before massive stars evolve into compact objects \citep{pzwart2002, pzwart2004, gaburov2008}. Repeated stellar collisions have been shown to be effective in clusters with masses $>10^5\,\msun$ and central densities $>5\times10^6\,\msun\,\rm pc^{-3}$ \citep{rizzuto2021, gonzalez2021, mas2023, rantala2024, rantala2025,
rantala2026}, corresponding to structural parameters that are rarely observed in local YMCs. Whether stellar collisions can efficiently promote the growth of VMSs in YMCs more commonly found in the local Universe therefore remains uncertain, partly because of the computational challenge of simulating massive
clusters with high primordial binary fractions (i.e., the fraction of binaries born with the cluster).

The dynamical evolution of YMCs has been investigated using a variety of numerical techniques, including Monte Carlo methods \citep[e.g.,][]{mocca2015, rodriguez2022}, direct $N$-body simulations \citep[e.g.,][]{wang2016, dicarlo2020, rastello2021, mas2024}, and hybrid $N$-body approaches \citep[e.g.,][]{wang2020b, barber2025}. However, most $N$-body studies have focused either on extremely dense clusters with relatively low total masses, adopted low primordial binary fractions, or relied on simplified prescriptions for stellar evolution. As a consequence, the combined impact of the high primordial binary fractions inferred for massive stars in local associations \citep{sana2012, moe2017} and repeated stellar collisions has
not yet been systematically explored in YMC models employing up-to-date single and binary stellar evolution.

In this work, we present the \textsc{titans} simulation set, a new suite of $N$-body simulations performed with the state-of-the-art $N$-body code \petar\footnote{\url{https://github.com/lwang-astro/PeTar}}. Our models span wide ranges of cluster masses  ($8\times10^{4}-9\times10^{5}\,\msun$) and half-mass densities ($100-10^5\,\msun\,\rm pc^{-3}$) adopting observation-based primordial binary fractions \citep{moe2017}. While the ultimate goal is to compare the properties of these clusters to those of local globular clusters, we focus here on the first $20\,\rm Myr$ of cluster evolution (\emph{Teen} \textsc{TITANS}), corresponding to the phase in which massive stars are still evolving into compact objects or pair-instability supernovae (PISNe). Using realistic initial conditions for YMCs \citep{pzwart2010, bastian2013, krumholz2019} and stellar-evolution prescriptions that include pair-instability processes, we self-consistently study the efficiency of repeated stellar collisions, the formation and evolution of VMSs, and their impact on the BH mass spectrum in YMCs.

The paper is structured as follows. Section~\ref{sec:methods} presents the code used and the chosen initial conditions. Section~\ref{sec:results} reports the main results on repeated stellar collisions and their final stellar products. Section~\ref{sec:discussion} discusses the impact of our results on the formation of PISNe and on the BH mass distributions in our clusters. Finally, Section~\ref{sec:summary} summarizes our findings.

\section{Methods}\label{sec:methods}
The \textsc{titans} set comprises 18 $N$-body simulations of star clusters in a Milky Way-like galaxy, taking into account single and binary stellar evolution, and considering a set of structural parameters compatible with observational limits in local YMCs. We perform this set of simulations with the hybrid $N$-body code \petar \citep{wang2020b}. Hybrid $N$-body codes are based on Hamiltonian splitting. In particular, \petar employs the particle-tree particle-particle algorithm (P$^3$T; \citealp{oshino2011}), and the slow-down algorithmic regularization method (SDAR; \citealp{wang2020a}) for an accurate and efficient treatment of binary systems and close encounters. The use of OpenMP for an efficient workload distribution across central processing units (CPUs), combined with the graphic processing unit (GPU) acceleration, contributes significantly to the strong parallel performance and scalability of \petar. 
This code can include single and binary star evolution through population-synthesis algorithms. Here, we use the \mobse population-synthesis code \citep{mapelli2017,giacobbo2018}. \petar can also model the effect of an external potential with \galpy \citep{bovy2015}.

We have performed our simulations on two servers: \texttt{demoblack}, hosted by the University of Padova, with 8 Nvidia Tesla V100 GPUs and 8 INTEL Xeon Platinum CPUs with 24 cores each; and \texttt{bwForCluster Helix} on a partition hosting 4-8 Nvidia A100 GPUs and 2 AMD EPYC CPUs with 64 cores each.  

\subsection{Stellar and binary evolution}
In our set of simulations, we have modeled single and binary stellar evolution using the population synthesis code \mobse \citep{giacobbo2018, giacobbo2020}. All clusters have metallicity $Z = 0.0002$, a value typical of metal-poor globular clusters \citep{vandenberg2013}, and favoring the formation of massive compact remnants \citep{askar2017,rantala2024,mas2024,barber2025}. 

Binary evolution processes, such as tides, mass transfer, common-envelope evolution, and GW-driven orbital decay, are implemented following the prescriptions of \cite{hurley2002}. In this work, we model the common-envelope phase using the standard $\alpha_{\rm CE}$–$\lambda_{\rm CE}$ formalism, with $\alpha_{\rm CE}$ the fraction of orbital energy used to unbind the envelope, and $\lambda_{\rm CE}$ a parameter characterizing the envelope's binding energy and structure. In particular, we adopt $\alpha_{\rm CE} = 1$, while $\lambda_{\rm CE}$ is calculated according to \cite{claeys2014}.

For core-collapse supernovae (SNe), we rely on the delayed formalism by \cite{fryer2012}. This model adopts a convection-enhanced neutrino-driven mechanism for the SN explosion, and the revival of the shock wave happens $\sim 250\,\rm ms$ after the collapse. With this prescription the minimum BH mass is $3\,\msun$. 

After the SN explosion, compact remnants receive a natal kick modeled following \cite{giacobbo2020}:
\begin{equation}
     v_{\rm kick} = f_{\rm H05}\,\frac{\langle m_{\rm NS} \rangle}{m_{\rm rem}}\,\frac{m_{\rm ej}}{\langle m_{\rm ej} \rangle},
\end{equation}
where $\langle m_{\rm NS} \rangle$ and $\langle m_{\rm ej} \rangle$ are the average neutron-star mass and ejecta mass, computed from a population of isolated neutron stars with metallicities representative of the Milky Way, $m_{\rm rem}$ is the mass of the compact object, and $m_{\rm ej}$ is the ejected mass. $f_{\rm H05}$ is a number randomly drawn from a Maxwellian distribution with one-dimensional root mean square $\sigma_{\rm kick} = 265\,\mathrm{km\,s^{-1}}$, derived from the proper motions of young Galactic pulsars \citep{hobbs2005}. This formalism implies that BHs forming through direct collapse receive a null natal kick. In this study we do not include a formalism for the natal spin of BHs.

For PISNe and pulsational PISNe we have used the fitting formulas reported in the appendix of   \cite{mapelli2020}, based on \cite{woosley2017}. When the helium core mass of a star, $m_{\rm He,f}$, is between 32 and $64\,\msun$, it undergoes a pulsational PISN. A PISN is triggered instead when $64\le m_{\rm He,f}\le 135\,\msun$. For isolated single stellar evolution at the metallicity adopted in our simulations, this corresponds to a range in the zero-age main-sequence mass $140\lesssim m_{\rm ZAMS}\lesssim 260\,\msun$. When $m_{\rm He,f}>135\,\msun$, the star directly collapses into a BH.

\subsection{Compact object mergers in \petar}
In dense star clusters with a high fraction of primordial binaries, collisions and mergers between stars and BHs can happen frequently (see e.g. \citealp{rastello2025}). The outcome of such interactions is still unclear. On the one hand, we expect that stars with $m_*<10\,\msun$ would accrete little or no mass on the BH \citep{kremer2020}. On the other, simulations from \cite{schroder2020} show that common envelope events between a star and a BH can lead to the accretion of the stellar core on the BH and to the ejection of the envelope. Moreover, various dynamical studies \citep{kiroglu2025, kiroglu2025b} have found that BHs can accrete mass from repeated collisions with stars in a dense cluster. As a consequence, previous works \citep{banerjee2021, rizzuto2021, mas2024} have assumed a fraction of stellar accreted mass on the BH $f_{\rm c}>0$. In this work, we have adopted a more conservative approach, assuming $f_{\rm c}=0$.

In \petar, GW–driven mergers are modeled by evolving the semi-major axis and eccentricity of the binary according to the formalism of \cite{peters1964}. A binary is considered to merge when its GW timescale becomes shorter than its integration timestep, which depends on the strength of nearby perturbations and the binary’s slowdown factor. Once this condition is met, \petar evolves the binary’s position to the merger time, allowing us to locate binary BH (BBH) mergers within the cluster. 

\subsection{External potential}
We model the influence of an external galactic potential using the \galpy package \citep{bovy2015}. Specifically, we assume that the star clusters reside in a Milky Way–like galaxy and follow a circular orbit around the Galactic center at a radius of $8\,\rm kpc$ with a velocity of $220\,\rm km\,s^{-1}$, corresponding to the Solar neighborhood. For the galactic potential, we adopt the \texttt{MWPotential2014} model, which includes a Miyamoto-Nagai potential for the disk \citep{miyamoto1975}, a spherical bulge component based on a truncated power-law density profile \citep{hernquist1990}, and a Navarro-Frenk-White potential for the dark matter halo \citep{nfw1996}.

\subsection{Initial conditions}
The initial conditions of our simulations are generated using \textsc{mcluster} \citep{kuepper2011}. We adopt initial cluster masses in the range $M_{\rm cl} \in \left[8\times10^4, 9\times10^5\right]\,\msun$, with positions and velocities sampled from a King density profile \citep{king1966} characterized by a central potential parameter $W_0 = 6$ and half-mass radii $r_{\rm h} \in \left[1, 5\right]\,\rm pc$. These values are consistent with those measured for YMCs in the local Universe \citep{pzwart2010, bastian2013, krumholz2019}. The initial density at half-mass radius $\rho_{\rm h}$ of our clusters is between $100$ and $10^5\,\msun\,\rm pc^{-3}$ and the half-mass relaxation time $t_{\rm rh}$ \citep{spitzer1988} is between $100\,\rm Myr$ and $3\,\rm Gyr$. A summary of the initial conditions for our suite of simulations is provided in Table~\ref{table:ic_sims}. 

Stellar masses are sampled from a Kroupa initial mass function \citep{kroupa2001} between $0.08$ and $150\,\msun$. We assume that the primordial binary fraction depends on the primary mass of the star, according to the prescriptions in \cite{moe2017}. The global binary fraction $f_{\rm b}$ is defined as $f_{\rm b} = N_{\rm b}/(N_{\rm s} + N_{\rm b})$ and its value is $\sim0.23$ across all the models, corresponding to $\sim38\%$ of stars in a primordial binary. Binaries are initialized with distributions of mass ratio, semi-major axis, and eccentricity motivated by the intrinsic properties of massive binaries inferred by \cite{sana2012}. We extend the orbital period range up to $10^5\,\rm days$ to include wide, non-interacting binaries. In addition, the mass-ratio distribution is extrapolated to $q<0.1$ to allow for extreme mass-ratio systems.

\begin{figure}[ht]
    \centering
    \includegraphics[width=0.9\columnwidth]{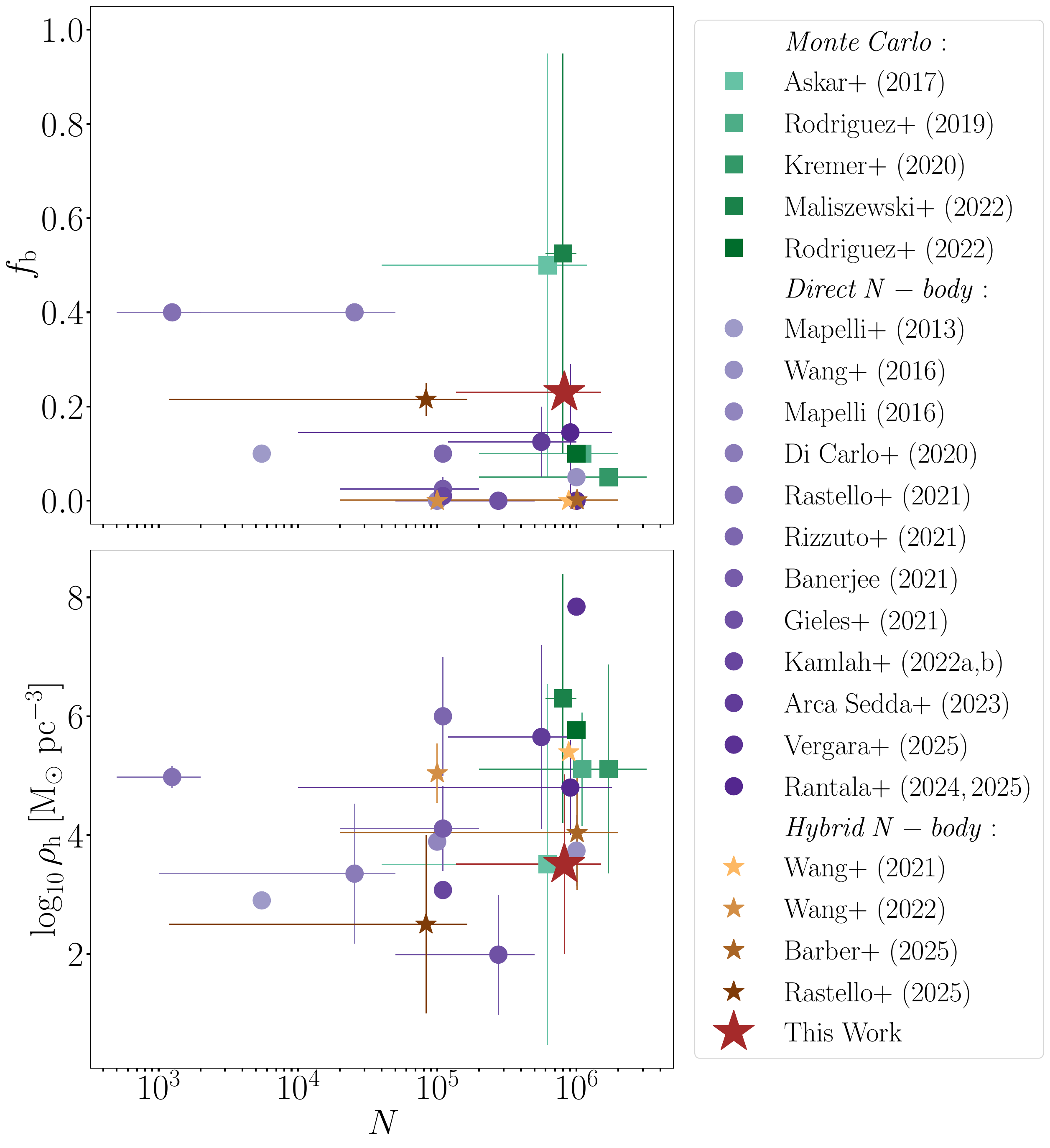}
    \caption{Initial binary fraction (top row) and density at half-mass radius (bottom row) as a function of the initial number of stars, for Monte Carlo (green squares), direct $N$-body (violet circles) and hybrid $N$-body simulations (brown stars). The \textsc{titans} simulations are represented by a red star.}
     \label{fig:init_cond}
\end{figure}

Figure~\ref{fig:init_cond} shows the location of our simulations in the $N-\rho_{\rm h}$ and $N-f_{\rm b}$ planes, where $N$ is the number of particles. We compare our models with state-of-the-art studies performed using Monte Carlo methods (e.g., \citealp{askar2017, rodriguez2019, maliszewski2022}), direct $N$-body simulations (e.g., \citealp{mapelli2013, mapelli2016, wang2016, mas2024, rantala2024}), and hybrid $N$ body approaches (e.g., \citealp{wang2021, rastello2025}). The main novelty of the \textsc{titans} simulation set lies in the fact that, although it explores a region of the $N-\rho_{\rm h}$ parameter space already covered by previous studies, it does so adopting a global primordial binary fraction $f_{\rm b}$ that has so far been investigated almost exclusively using Monte Carlo techniques \citep{askar2017, maliszewski2022}. Importantly, our simulations sample a region of parameter space that is representative of YMCs and globular clusters in the local Universe, whose typical densities are $\lesssim10^6\,\msun\,\rm pc^{-3}$.

The total simulated time for each cluster, $t_{\rm sim}$, exceeds $20\,\rm Myr$ in all cases. For the first $20\,\rm Myr$, the evolution of each star cluster is tracked in greater detail, with output snapshots recorded every $0.125\,\rm Myr$. 
In this work, we focus on the first $20\,\rm Myr$ of cluster evolution, corresponding to the lifetime of a $\sim12\,\msun$ star; hence, we expect that all BHs have formed by this time and most SNe have already taken place. Moreover, within $20\,\rm Myr$, the densest clusters contract and undergo core collapse \citep{spitzer1988}. 
After $t_{\rm sim} > 20\,\rm Myr$, output snapshots are recorded every $1\,\rm Myr$.

\begin{table*}[ht!]
\centering
\caption{Initial conditions of the \textsc{titans} set}
\begin{tabular}{c c c c c c c c c} 
 \hline
 Model & $N$ & $M_{\rm cl}$ & $N_{\rm b}$ & $r_{\rm h}$ & $\rho_{\rm h}$ & $t_{\rm rh}$ & $t_{\rm sim}/t_{\rm rh}$ & $t_{\rm sim}/t_{\rm cc}$\\ [0.5ex] 
& $10^3$ & $10^5\,\msun$ & $10^3$ & pc & $10^3\,\msun\,\rm pc^{-3}$ & Gyr & &\\ [0.5ex] 
 \hline\hline 
 T1 & 136.9 & 0.8 & 25.8 & 1 & 9.7 & 0.1 & 51 & 255\\
 T2 & 195.4 & 1.2 & 36.7 & 5 & 0.1 & 1.3 & 4.3 & 21\\
 T3 & 372.9 & 2.2 & 70.3 & 2 & 3.3 & 0.4 & 4.7 & 23.7\\
 T4 & 407 & 2.4 & 76.7 & 2 & 3.7 & 0.4 & 0.05 & 0.3\\
 T5 & 411.1 & 2.5 & 77.5 & 2 & 3.7 & 0.4 & 0.7 & 3.3\\
 T6 & 528.4 & 3.2 & 99.7 & 5 & 0.3 & 1.9 & 0.3 & 1.3\\
 T7 & 600.8 & 3.6 & 113.3 & 5 & 0.4 & 2 & 1 & 5.3\\
 T8 & 825.7 & 4.9 & 155.6 & 5 & 0.5 & 2.3 & 0.1 & 0.5\\
 T9 & {823.2} & {5} & {155.3} & {1} & {59.1} & {0.2} & {0.1} & 0.5\\
 T10 & 850.7 & 5.1 & 160.4 & 5 & 0.5 & 2.3 & 0.3 & 2\\
 T11 & 1000 & 6.1 & 190.3 & 1 & 9.1 & 0.6 & 0.4 & 2\\
 T12 & 1230.6 & 7.4 & 232 & 2 & 11.1 & 0.7 & 0.1 & 0.5\\
 T13 & 1262.1 & 7.6 & 237.8 & 5 & 0.7 & 2.7 & 0.02 & 0.1\\
 T14 & {1314.2} & {8} & {247.9} & {2} & {11.9} & {0.7} & {0.03} & 0.1\\
 T15 & 1343.2 & 8.1 & 253.1 & 2 & 12 & 0.7 & 0.03 & 0.1\\
 T16 & 1430.8 & 8.6 & 270 & 1 & 102.7 & 0.3 & 0.07 & 0.3\\
 T17 & 1461.4 & 8.8 & 275.3 & 5 & 0.8 & 2.9 & 0.1 & 0.6\\
 T18 & 1508.7 & 9.1 & 284.5 & 5 & 0.9 & 2.9 & 0.02 & 0.1\\
 \hline
\end{tabular}
\tablefoot{Column 1: name of the model. Column 2: number of stars $N = N_{\rm s}+2\,N_{\rm b}$. Column 3: cluster mass. Column 4: number of binary systems. Column 5: half-mass radius of the cluster. Column 6: density at half-mass radius of the cluster. Column 7: half-mass relaxation time \protect\citep{spitzer1988}. Column 8 and 9: fraction of simulated time with respect to the relaxation time $t_{\rm rh}$ and the core-collapse time $t_{\rm cc}\sim0.2\,t_{\rm rh}$.}
\label{table:ic_sims}
\end{table*}

\subsection{Properties of stellar collisions}\label{methods:stell_coll}

Stellar collisions and repeated stellar collisions have been extensively investigated as formation channels of IMBH progenitors \citep{pzwart2002, pzwart2004, mocca2015,mapelli2016, dicarlo2021, gonzalez2021, rizzuto2021, kritos2023, reinoso2023, mas2024b, rantala2024, rantala2025, vergara2025, paiella2025}. 
In this work, we explore the properties of repeated stellar collisions in YMCs with a high $f_{\rm b}$, and features typical of local star clusters.

\begin{figure}[ht]
    \centering
    \includegraphics[width=\columnwidth]{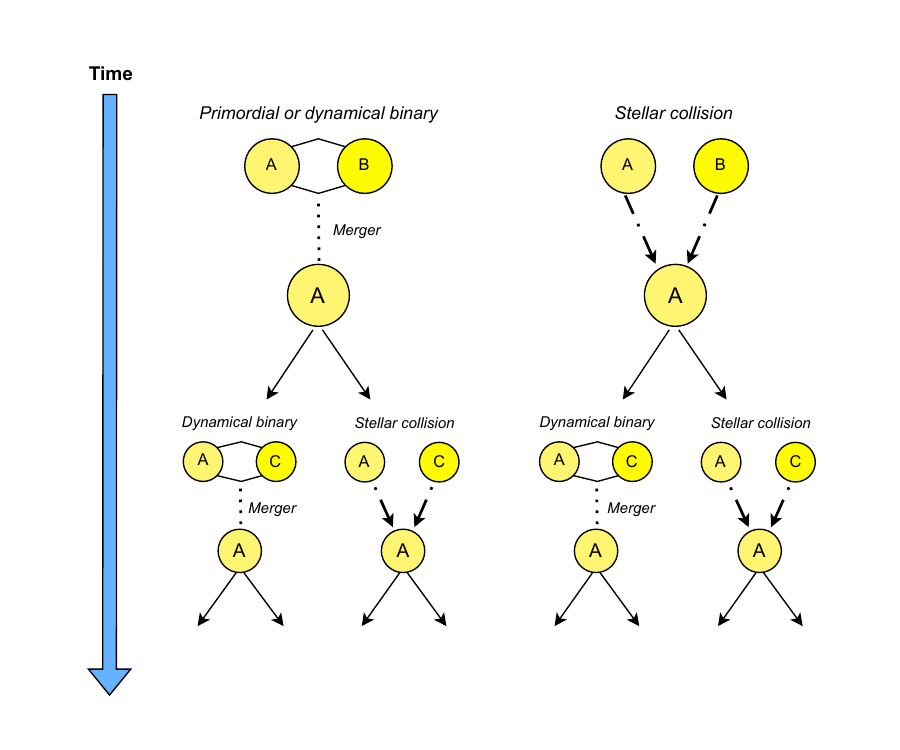}
    \caption{Schematic representation of a stellar collision chain. We show on the left a chain initiated by the merger of the components of a binary (either primordial or dynamically formed). On the right we show instead a chain initiated by an hyperbolic stellar collision.}
     \label{fig:coll_chain_diagram}
\end{figure}

Here, we define a stellar collision chain as a sequence containing at least two stellar collisions or mergers. Throughout the paper, we use the term “stellar collision” in a broad sense, referring both to mergers of binary components and to direct collisions between unbound stars. We quantify the efficiency of repeated stellar collisions as $\eta_{\rm rep} = N_{\rm ch,rep}/M_{\rm cl}$, where $N_{\rm ch,rep}$ is the number of stellar collision chains identified in each cluster.  Figure~\ref{fig:coll_chain_diagram} represents schematically a stellar collision chain, initiated either by the merger of a primordial or dynamical binary (i.e. a binary that is not born with the cluster), or by a stellar collision. The resulting star will itself either form a dynamical binary that will later merge, or collide with another star. We expect this process to re-iterate itself for several times, especially in dense clusters with small relaxation times. When the remnants from two or more chains merge, we compute $N_{\rm ch,rep}$ as the sum of the length of the chains.

We also examine how repeated stellar collisions affect the mass spectrum of VMSs and IMBHs. All through this study, we define VMSs as stars whose mass exceeds the upper limit of the adopted initial mass function ($m_* > 150\,\msun$), and IMBHs as BHs with $m_{\rm BH}>100\,\msun$.

\section{Results}\label{sec:results}
In the following Sections, we discuss the properties of repeated stellar collisions (Sec.~\ref{sec:rep_coll_gen}), and their impact on the formation of VMSs (Sec.~\ref{sec:vms}). We will show our results considering only the first $20\,\rm Myr$ of evolution of each cluster. 

\subsection{Repeated stellar collisions}\label{sec:rep_coll_gen}
\subsubsection{Impact of the host environment}\label{sec:repeated_colls}
\begin{figure}[ht]
    \centering
    \includegraphics[width=0.9\columnwidth]{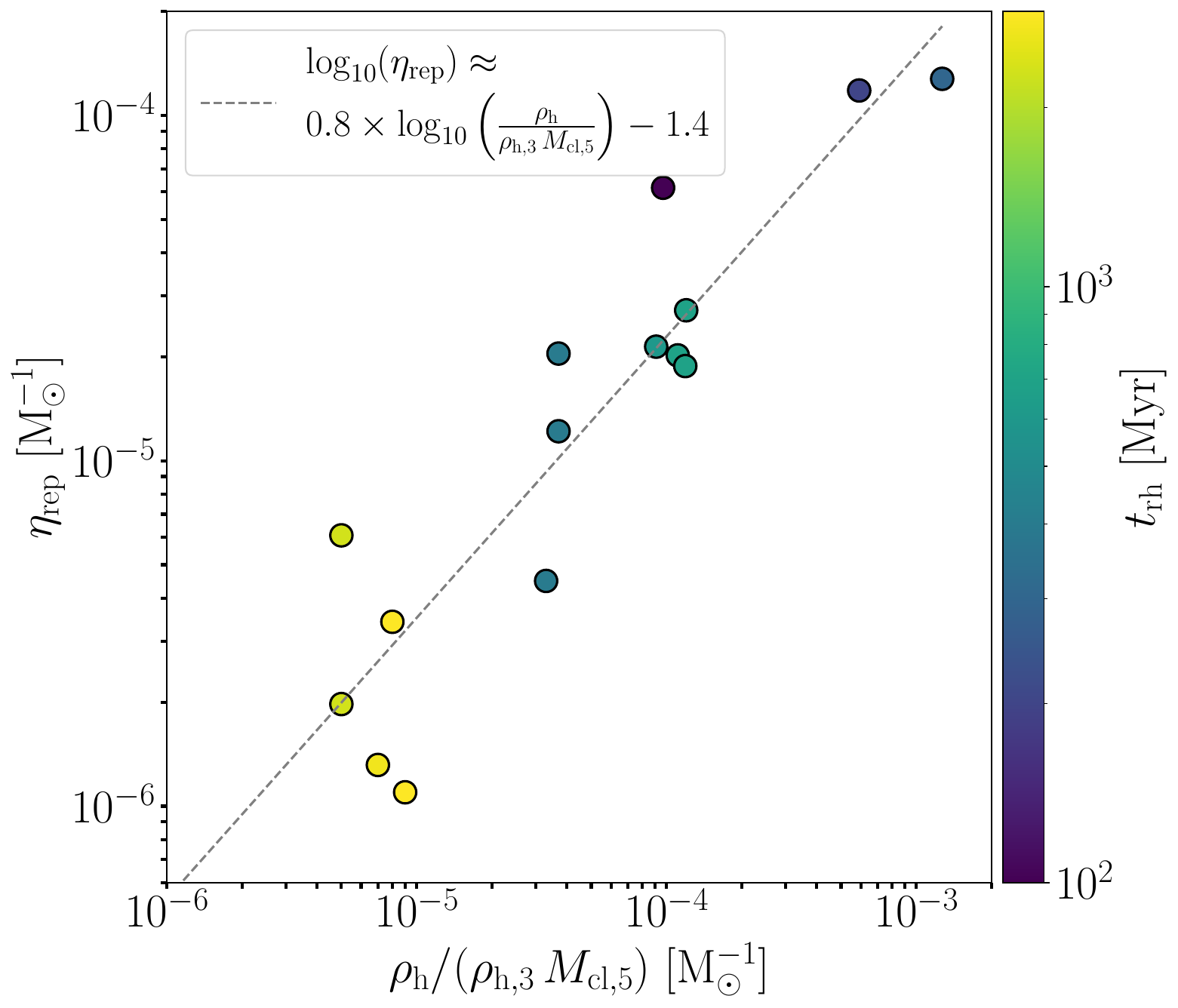}
    \caption{Efficiency of repeated stellar collisions $\eta_{\rm rep}$ as a function of the half-mass density $\rho_{\rm h}$, normalized by $\rho_{\rm h,3}$ and $M_{\rm cl,5}$. The points are colored as a function of the cluster initial relaxation time $t_{\rm rh}$.}
     \label{fig:eff_rhoh}
\end{figure}

Figure~\ref{fig:eff_rhoh} shows the efficiency of repeated stellar collisions as a function of $\rho_{\rm h}/(\rho_{\rm h,3}\,M_{\rm cl,5})$, with $\rho_{\rm h,3}=10^3\,\msun\,\rm pc^{-3}$ and $M_{\rm cl,5}=10^5\,\msun$. As defined in Sec.~\ref{methods:stell_coll}, the efficiency of repeated collisions is $\eta_{\rm rep}=N_{\rm ch,rep}/M_{\rm cl}$, where $N_{\rm ch,rep}$ is the number of stellar collision chains identified in each cluster. As cluster density increases, so does the efficiency of repeated collisions. We fit the data with a linear relation in logarithmic space, obtaining
\begin{equation}
    \log_{10}(\eta_{\rm rep})\approx
    0.8\,\log_{10}\!\left(\frac{\rho_{\rm h}}{\rho_{\rm h,3}\,M_{\rm cl,5}}\right)-1.4
    \,\,\,\rm with\,\,\,\rho_{\rm h}\gtrsim500\,\msun\,\rm pc^{-3}.
\end{equation}

 The intrinsic scatter around the best-fit relation is $\sigma_{\rm int}\sim0.3\,\rm dex$. This trend indicates that a high cluster density is a key requirement for the onset of multiple stellar collisions. Moreover, we find that repeated collisions occur only for $\rho_{\rm h}\gtrsim500\,\msun\,\rm pc^{-3}$.

An opposite trend is observed with respect to the relaxation time $t_{\rm rh}$. At fixed density, clusters with shorter relaxation times exhibit a higher efficiency of repeated collisions. In fact, an efficient formation of merger chains requires a mass-segregation time $t_{\rm DF}\sim\langle m\rangle\,t_{\rm rh}/m_{\rm bin}$ shorter than the lifetime of the stars involved, $t_{\rm DF}<t_{\rm end}$, where the average stellar mass is $\langle m\rangle\sim0.6\,\msun$ and the binary mass satisfies $m_{\rm bin}\leq300\,\msun$. For the most massive binaries in our clusters ($m_{\rm bin}=300\,\msun$, $m_1=m_2=150\,\msun$), the stellar lifetime is $t_{\rm end}\sim3\,\rm Myr$, corresponding to a maximum relaxation time $t_{\rm rh}\sim1.5\,\rm Gyr$. Less massive binaries will require longer times to segregate to the cluster center.

As shown in Table~\ref{table:ic_sims}, more than half of the \textsc{titans} models have $t_{\rm rh}\lesssim1.5\,\rm Gyr$. As a consequence, despite their relatively long core-collapse times ($t_{\rm cc}>20\,\rm Myr$), these clusters can form stellar collision chains because the most massive primordial binaries segregate rapidly to the center. Once in the core, these binaries merge, and the increasing central density of massive stars favors the formation of new dynamical binaries or direct stellar collisions. This result is consistent with the findings of \cite{rantala2025} on the key role of primordial binaries in clusters with long core-collapse times. In contrast, clusters with $t_{\rm rh}\gtrsim1.5\,\rm Gyr$ require more time for binaries to segregate and interact; by the time they reach the core, the most massive stars are often already evolved or have died as compact objects, reducing the efficiency of chain formation.

\begin{figure}[ht]
    \centering
    \includegraphics[width=0.9\columnwidth]{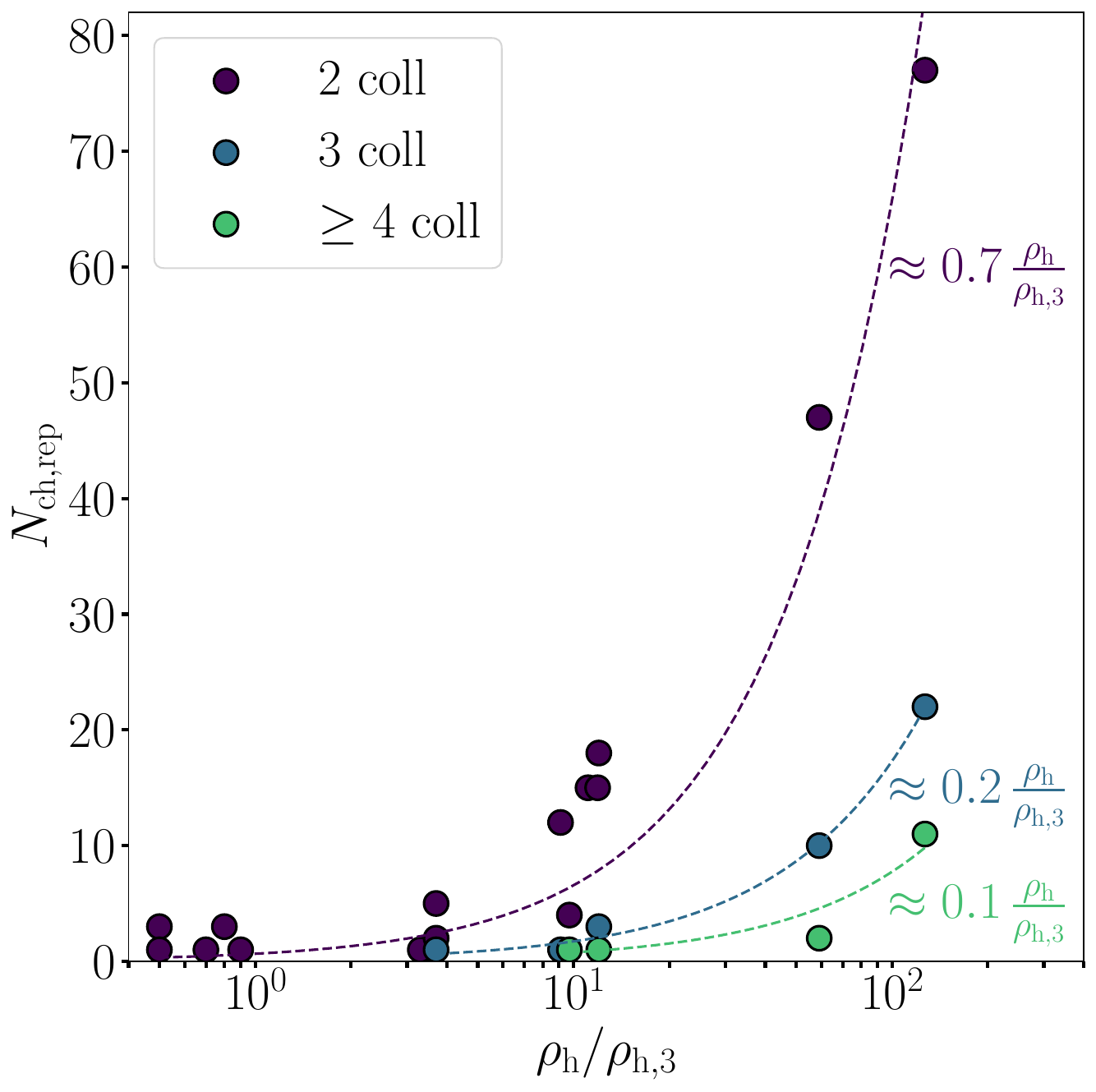}
    \caption{Number of stellar collision chains $N_{\rm ch,rep}$ as a function of the normalized density at half-mass radius $\rho_{\rm h}/\rho_{\rm h,3}$. In violet we represent the number of chains containing two collisions, in blue the number of chains containing three collisions and in four the number of chains containing four or more collisions. The dashed lines and annotations represent the linear fits of the points.}
     \label{fig:num_coll_chains}
\end{figure}

Figure~\ref{fig:num_coll_chains} shows the number of stellar collision chains $N_{\rm ch,rep}$ as a function of the normalized half-mass density $\rho_{\rm h}/\rho_{\rm h,3}$ and of the length of the chains themselves. The minimum density required for a chain to form increases with the number of collisions it contains: chains with two collisions appear at $\rho_{\rm h}\sim500\,\msun\,\rm pc^{-3}$, chains with three collisions require densities of $\sim4000\,\msun\,\rm pc^{-3}$ and chains with four or more collisions form only at densities above $\sim10^4\,\msun\,\rm pc^{-3}$. In all cases, $N_{\rm ch,rep}$ increases approximately linearly with density. However, the slope decreases for longer chains, indicating that extended collision sequences are rare  even in clusters with high densities and low relaxation times. As a result, the contribution of long collision chains to $N_{\rm ch,rep}$ and $\eta_{\rm rep}$ remains limited. In the densest model of our suite (T16, $\rho_{\rm h}\sim10^5\,\msun\,\rm pc^{-3}$), $70\%$ of the chains contain two collisions, $20\%$ contain three collisions, and only $10\%$ involve more than four collisions. We point out that models T2, T6, and T7, do not produce any stellar collision chains. Models T3, T10, T13, and T18 instead produce a single chain each ($N_{\rm ch,rep}=1$), always containing two collisions.

Finally, we compute the fraction of stellar collision chains whose first merger involves the components of a primordial binary. This fraction is $100\%$ up to $\rho_{\rm h}\sim10^4\,\msun\,\rm pc^{-3}$ and remains above $\sim72\%$ at higher densities. When the first collision does not involve a primordial binary merger, it results either from a hyperbolic encounter or from a highly eccentric collision between a low-mass main-sequence (MS) star and one component of a primordial binary. In the latter case, the perturber can be either a single star or a component of a low-mass primordial binary, which is disrupted during the interaction. Owing to the small mass ratio between the perturber and the massive primordial binary component, this first merger does not significantly alter the orbit of the primordial binary, whose components subsequently merge. 

Overall, these results highlight that stellar collision chains are efficient only in clusters with $\rho_{\rm h}\gtrsim10^4\,\msun\,\rm pc^{-3}$ and with $t_{\rm rh}\lesssim 300\,\rm Myr$ and that, even then, they mostly contain two collisions and are triggered by a primordial binary merger. In Secs.~\ref{sec:vms} and \ref{sec:imbh} we will explore the effect of these findings on VMSs and IMBHs.

\subsubsection{Stellar properties}\label{sec:st_types}

\begin{figure*}[ht]
    \centering
    \includegraphics[width=0.9\textwidth]{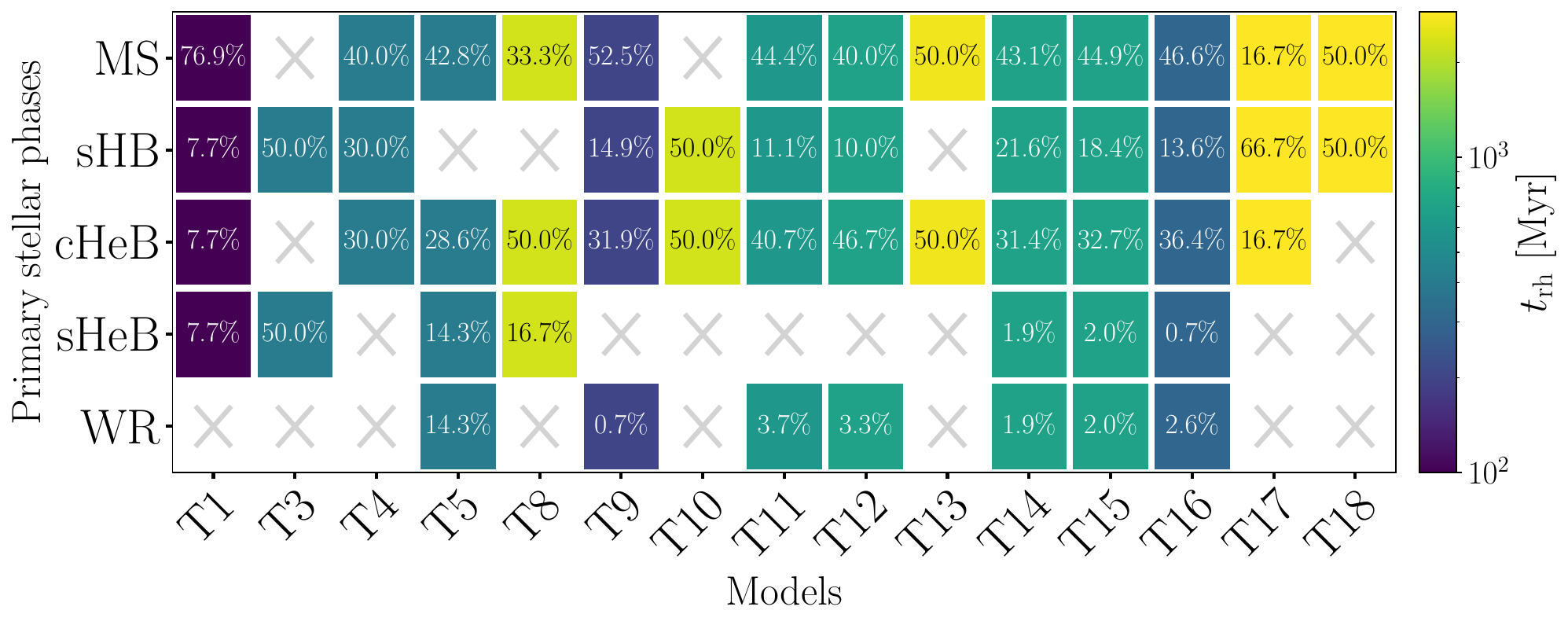}
    \caption{Primary stellar type of stars involved in repeated collisions. The $x$ axis shows the simulated models. The stellar phases on the $y$ axis are: MS, shell hydrogen burning (sHB), core helium burning (cHeB), shell helium burning (sHeB) and Wolf-Rayet star (WR). The color represents the relaxation time at half-mass radius $t_{\rm rh}$.}
     \label{fig:run_k1}
\end{figure*}

Figure~\ref{fig:run_k1} shows the stellar types of primary stars involved in repeated collisions in the simulated models, colored by the relaxation time of their host cluster. We do not show the stellar types of secondaries because in all models $\gtrsim 93\%$ stars are in their MS. Moreover, we do not show models T2, T6 and T7 since they do not produce stellar collision chains. In about half of the simulated models, the primary stars involved in repeated stellar collisions are preferentially in their MS. With increasing $t_{\rm rh}$ the fraction of evolved stars involved in repeated stellar collisions grows. In fact, longer relaxation times correspond to longer mass segregation timescales and, as a consequence, stars are more evolved by the time they reach the dense cluster center and take part in the collisions. Moreover, low densities lead to bigger intervals of time between consecutive collisions, allowing the stars to evolve. We move from model T1 ($t_{\rm rh}\sim100\,\rm Myr$) with $\sim77\%$ primaries in the MS, to model T17 ($t_{\rm rh}\sim3\,\rm Gyr$) with only $\sim17\%$ primaries in their MS. Models T3, T10 and T18, appearing as outliers, produce only one chain of stellar collisions. 

\begin{figure*}[ht]
    \centering
    \includegraphics[width=0.9\textwidth]{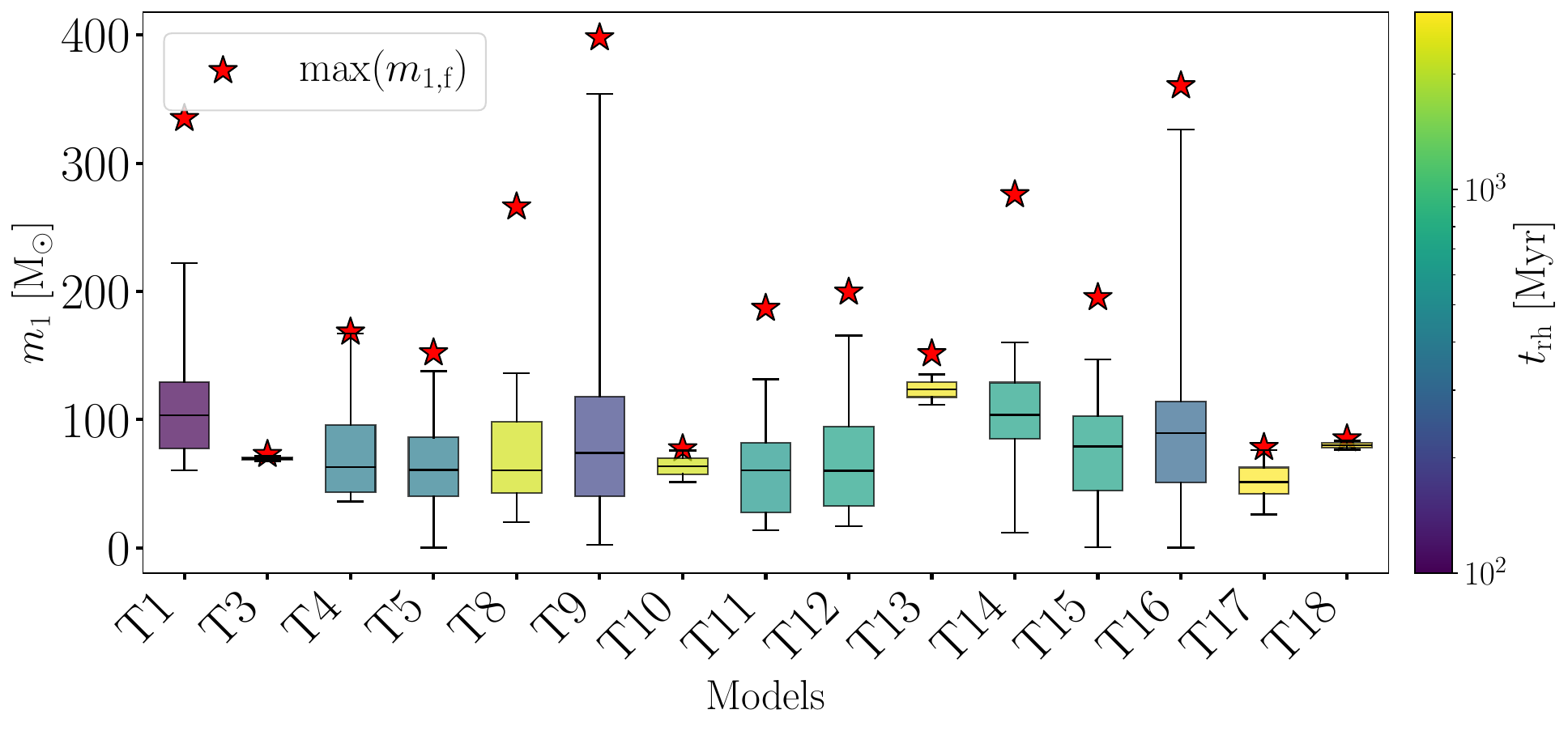}
    \caption{Distribution of primary mass $m_1$ of stars involved in repeated stellar collisions for each simulated model. The color represents the relaxation time at half-mass radius $t_{\rm rh}$. We show with a red star the maximum mass of the product of a stellar collision chain in each model.}
     \label{fig:run_m1}
\end{figure*}

Figure~\ref{fig:run_m1} shows the primary masses of all the stars involved in repeated stellar collisions. The bars contain all the mass values, while the rectangles contain $50\%$ of the data. We see that the bulk values of primary masses do not strongly depend on the properties of the host clusters. Dense clusters with $t_{\rm rh}\leq700\,\rm Myr$ (e.g. T1, T9, T14, T16) have larger bulk values ($50\lesssim m_1\lesssim 130\,\msun$) and they cover broader mass intervals. In fact, repeated stellar collisions are more frequent in dense clusters with low relaxation times (Fig.~\ref{fig:eff_rhoh}), and they mainly involve stars in their MS (Fig.~\ref{fig:run_k1}). Model T9 spans a primary mass interval extending up to $350\,\msun$, due to the long chains of repeated stellar collisions forming in this cluster. Models T13 and T18, appearing as outliers, have $N_{\rm ch,rep} = 1$. We also represent here the maximum mass reached in our clusters through stellar collision chains; we see that the highest values are reached in models T1, T9 and T16, which have the lowest relaxation times. For secondary masses, we find that the bulk values are usually below $m_2\lesssim 25\,\msun$ if $t_{\rm rh}\gtrsim500\,\rm Myr$. 

\subsection{Very massive stars}\label{sec:vms}
We summarize here the properties of the VMSs ($m_*>150\,\msun$) forming in our simulated clusters. Table~\ref{table:vms_info} lists, for each model, the number of VMSs, their maximum mass, and their formation channels. 

The number of VMSs generally increases with cluster mass, consistently with the fact that their dominant formation pathway is the merger of primordial binaries, whose abundance scales with $M_{\rm cl}$. In the densest clusters with short relaxation times, stellar collision chains play a non-negligible role in VMS formation. This is especially evident in models T1, T9, and T16, where more than $25\%$ of VMSs originate from repeated collisions. Models T9 and T16 form a larger number of VMSs than clusters of comparable mass because repeated stellar collisions are more efficient in these systems. In all the other models, the VMSs counted in $N_{\rm VMS,coll}$ form predominantly ($>82\%$) via primordial binary mergers. This result highlights again the key role of primordial binaries in driving the early dynamical and stellar evolution of YMCs with properties in line with low-redshift observations \citep{pzwart2010, rantala2025}. We find that binary evolution also contributes to VMS formation: in particular, stable mass transfer can produce stars with masses up to $\sim200\,\msun$. Across all clusters, the median VMS mass shows only a weak dependence on cluster properties and remains in the range $160-200\,\msun$. All VMSs identified in our simulations are still on the MS, as their formation typically occurs through stellar collisions involving young stars.

\begin{figure}[ht]
    \centering
    \includegraphics[width=0.9\columnwidth]{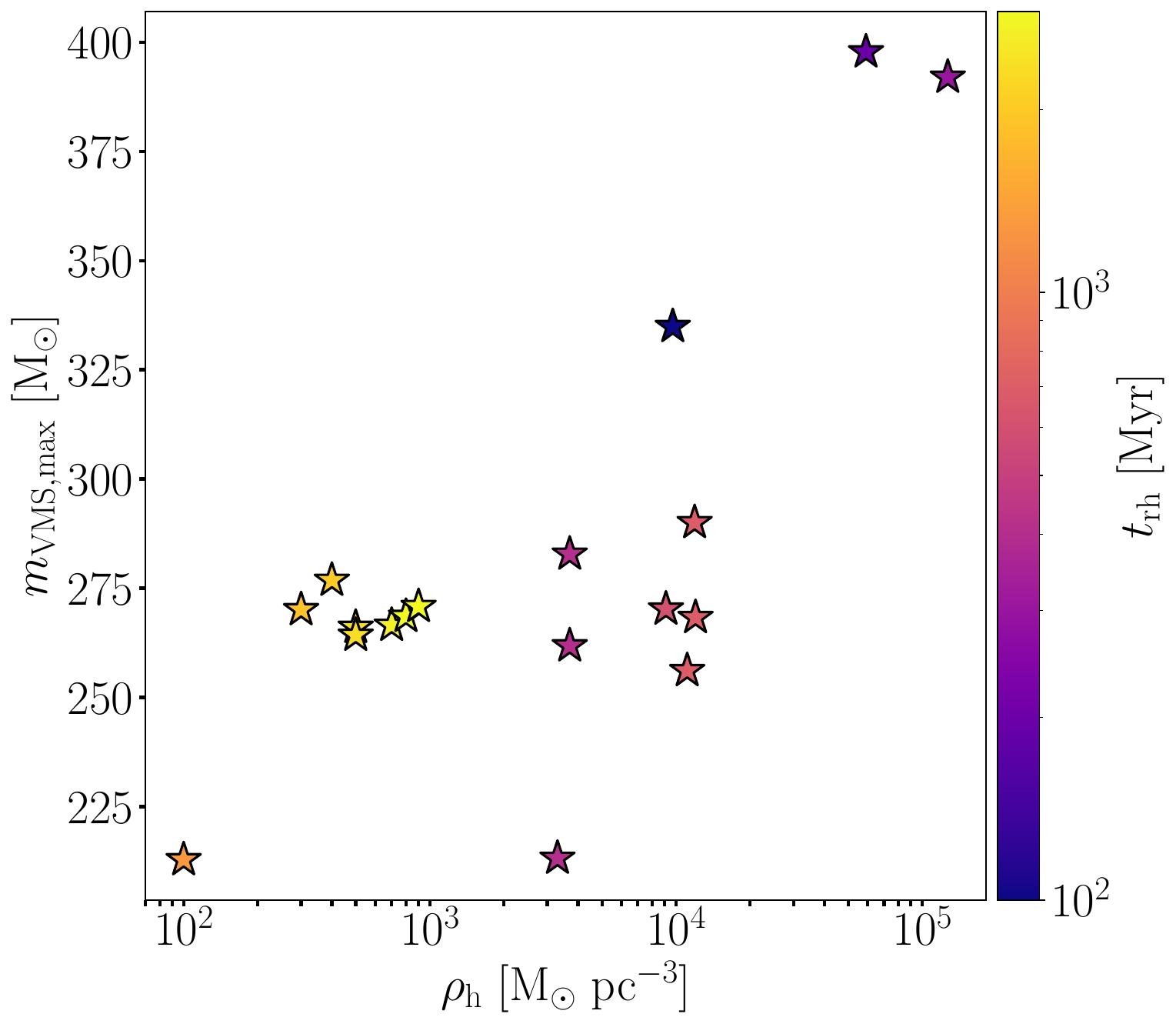}
    \caption{Relation between the maximum VMS mass $m_{\rm VMS,max}$ and the initial density at half-mass radius $\rho_{\rm h}$. The colors used represent the initial relaxation time $t_{\rm rh}$.}
     \label{fig:rho_max_vms}
\end{figure}

Table~\ref{table:vms_info} also reports the maximum stellar mass attained in each simulation. Figure~\ref{fig:rho_max_vms} shows it as a function of the initial half-mass density of the clusters. Since VMSs form early during cluster evolution ($t<5\,\rm Myr$), this trend remains unchanged when considering the density at the time of their formation. For clusters with $\rho_{\rm h}\lesssim10^4\,\msun\,\rm pc^{-3}$ and $t_{\rm rh}>300\,\rm Myr$, the maximum stellar mass is nearly independent of density and relaxation time, lying between $215$ and $300\,\msun$, with stochastic variations reflecting the collisional nature of VMS formation. In all these cases, VMSs originate from primordial binary mergers. In models with $\rho_{\rm h}\gtrsim10^4\,\msun\,\rm pc^{-3}$ and $t_{\rm rh}\lesssim300\,\rm Myr$, the maximum stellar mass exceeds $330\,\msun$. The three stars in this regime form in models T1, T9, and T16 through stellar collision chains.

Finally, the last column of Table~\ref{table:vms_info} reports the number of VMSs that merge with a stellar-mass BH. Appendix~\ref{app:fc_impact} discusses the final fate of these stars and the impact of our initial prescriptions. The subsequent evolution of VMSs is instead addressed in Sec.~\ref{sec:discussion}.

\begin{table*}[ht!]
\centering
\caption{Properties of VMSs in the simulated models}
\begin{tabular}{c c c c c c c} 
 \hline
 Model & $N_{\rm VMS}$ & $m_{\rm VMS,max}$ & $N_{\rm VMS,bin}$ & $N_{\rm VMS, coll}$ & $N_{\rm VMS,repcoll}$ & $N_{\rm BH, coll}$\\ [0.5ex] 
 &  &  $\msun$ & & & \\ [0.5ex] 
 \hline\hline
 T1 & 6 & 334.9 & 0 & 3 & 3 & 1\\
 T2 & 10 & 212.8 & 1 & 9 & 0 & 0 \\
 T3 & 13 & 213.2 & 1 & 12 & 0 & 1\\
 T4 & 15 & 261.8 & 3 & 11 & 1 & 0 \\
 T5 & 22 & 282.8 & 2 & 20 & 0 & 0\\
 T6 & 27 & 270.1 & 4 & 23 & 0 & 0\\
 T7 & 18 & 276.8 & 3 & 15 & 0 & 0\\
 T8 & 36 & 266 & 6 & 30 & 0 & 0\\
 T9 & 46 & 397.7 & 3 & 32 & 11 & 0\\
 T10 & 24 & 264.2 & 5 & 19 & 0 & 1\\
 T11 & 28 & 270.2 & 5 & 22 & 1 & 1\\
 T12 & 42 & 256.1 & 9 & 33 & 0 & 0\\
 T13 & 45 & 266.5 & 7 & 38 & 0 & 0\\
 T14 & 56 & 290 & 4 & 50 & 2 & 0\\
 T15 & 40 & 268.3 & 4 & 36 & 0 & 1\\
 T16 & 84 & 391.9 & 7 & 61 & 16 & 4\\
 T17 & 58 & 268.5 & 13 & 45 & 0 & 0\\
 T18 & 50 & 270.8 & 3 & 47 & 0 & 0\\
 \hline
\end{tabular}
\tablefoot{Column 1: name of the model. Column 2: total number of VMSs. Column 3: maximum mass of VMSs in each cluster. Column 4: number of VMSs due to binary evolution processes. Column 5: number of VMSs due to single collisions. Column 6: number of VMSs due to chains of collisions. Column 7: number of VMSs colliding with a stellar-mass BH.}
\label{table:vms_info}
\end{table*}

\section{Discussion}\label{sec:discussion}
In the following, we discuss the impact of stellar collisions and repeated stellar collisions on PISNe (Sec.~\ref{sec:pisn}), and the mass distribution of BHs (Sec.~\ref{sec:bh}). We focus particularly on the formation and properties of IMBHs in Sec.~\ref{sec:imbh}. Also here, we consider only the first $20\,\rm Myr$ of evolution of each cluster, unless otherwise specified. In Sec.~\ref{sec:caveats} we report the caveats and limitations of our models.  

\subsection{Pair-instability supernovae}\label{sec:pisn}

We expect that stellar collisions increase the number of stars entering the PISN mass range relative to predictions from the initial mass function, potentially enhancing the rates of PISNe and pulsational PISNe in metal-poor star clusters compared to the field.

\begin{figure}[ht]
    \centering
    \includegraphics[width=0.9\columnwidth]{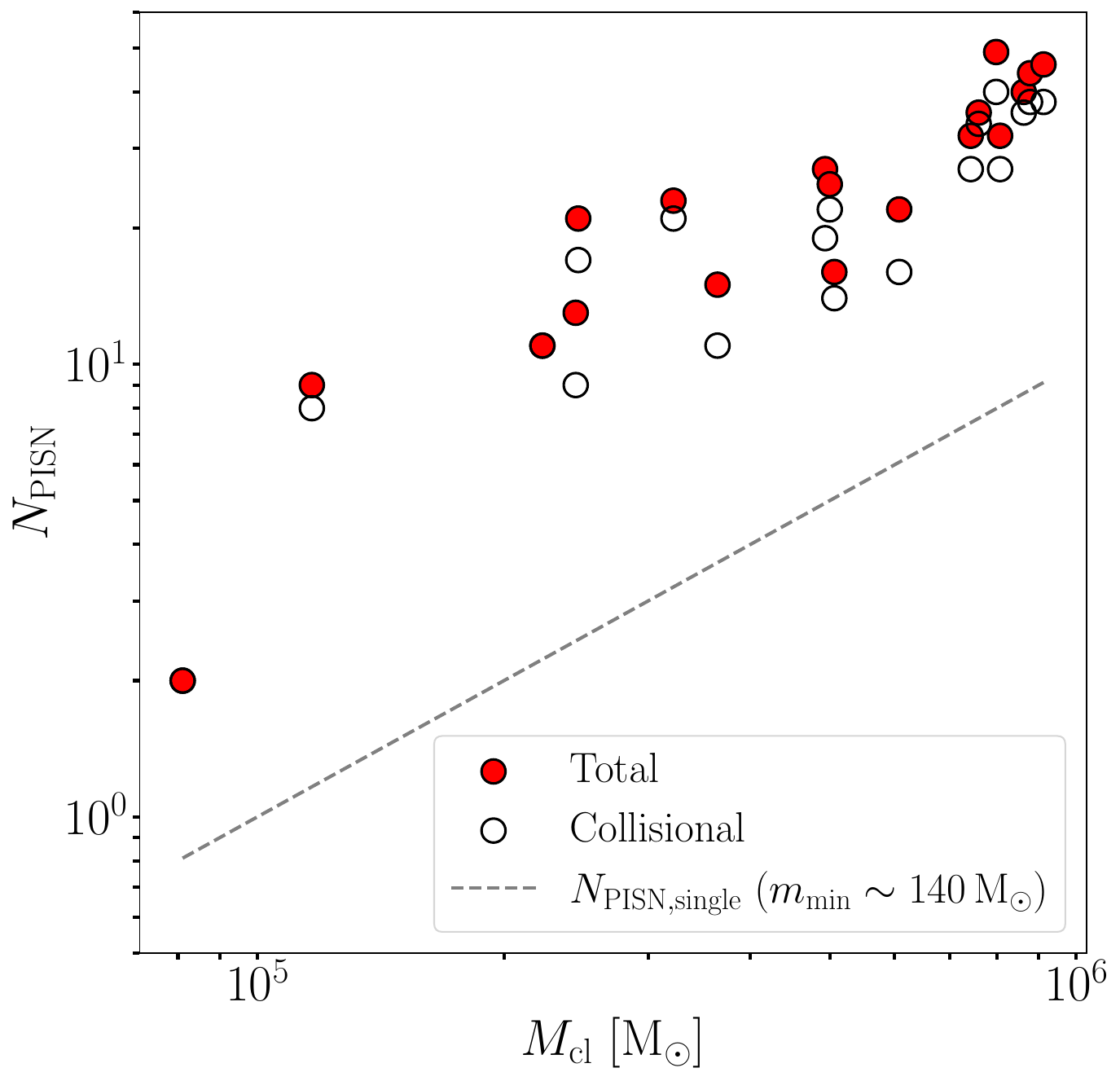}
    \caption{Number of PISNe as a function of the cluster mass $M_{\rm cl}$. The filled red points represent the total number; the empty black points represent instead the number of PISNe due to collisions. As a comparison we plot the number of PISNe expected from single stellar evolution $N_{\rm PISN,single}$.}
     \label{fig:num_pisn_mcl}
\end{figure}

In our simulations, most PISNe originate either from stars that are born directly within the PISN mass range or from stars that enter it following a primordial binary merger. Since none of the PISN progenitors escape from the clusters, both formation channels scale with the cluster mass $M_{\rm cl}$. This behavior is confirmed in Fig.~\ref{fig:num_pisn_mcl}, which shows that both the total number of PISNe and the fraction produced through stellar collisions correlate strongly with $M_{\rm cl}$. The contribution of stellar collisions to the total PISN population is above $\sim70\%$, with large scatter among clusters with similar properties, reflecting the stochastic nature of stellar collisions and massive binary evolution. Only four PISNe (3 in model T9, 1 in model T16) form via repeated stellar collisions. We find that $\gtrsim90\%$ of our PISNe explode as single stars, while the remaining part is found in a binary system. The only exception is model T1, where $50\%$ of the PISNe explode in a binary. This difference is driven by small-number statistics, as T1 is the least massive cluster in our sample and forms only two PISNe.

Figure~\ref{fig:num_pisn_mcl} also shows the expected number of PISNe from an isolated population of single stars, assuming that the PISN mass gap starts at $\sim140\,\msun$ \citep{spera2017, giacobbo2018b}. We find that the number of stars born in our clusters within the PISN mass range is consistent with the isolated case. However, stellar collisions (primarily in the form of primordial binary mergers) and, to a lesser extent, repeated stellar collisions, allow additional stars to enter the PISN regime, increasing the total number of PISNe. As a result, in most of our models the total number of PISNe forming in star clusters is up to one order of magnitude larger than expected for an isolated single stellar population.

In our simulations, the PISN mass gap corresponds to $140\lesssim m_*\lesssim400\,\msun$ considering the final stellar mass before the explosion, with the bulk of the population $<289\,\msun$ \citep[e.g.,][]{mas2024}. We expect these results to be robust against uncertainties in stellar-wind prescriptions owing to the low metallicity adopted in this work \citep{simonato2025}.

Finally, we find that $\gtrsim70\%$ of the VMSs formed in the clusters explode as PISNe. The remaining VMSs typically lose a substantial fraction of their mass and collapse into stellar-mass black holes with $m_{\rm BH}<100\,\msun$ as a result of pulsational PISNe (see Sec.~\ref{sec:bh}), or are removed through collisions with stellar-mass black holes (see Appendix~\ref{app:fc_impact}). The only exception is model T1, in which $\sim33\%$ of the VMSs die as PISNe.

\subsection{Black hole mass distribution}\label{sec:bh}

\begin{figure*}[ht]
    \centering
    \includegraphics[width=0.89\textwidth]{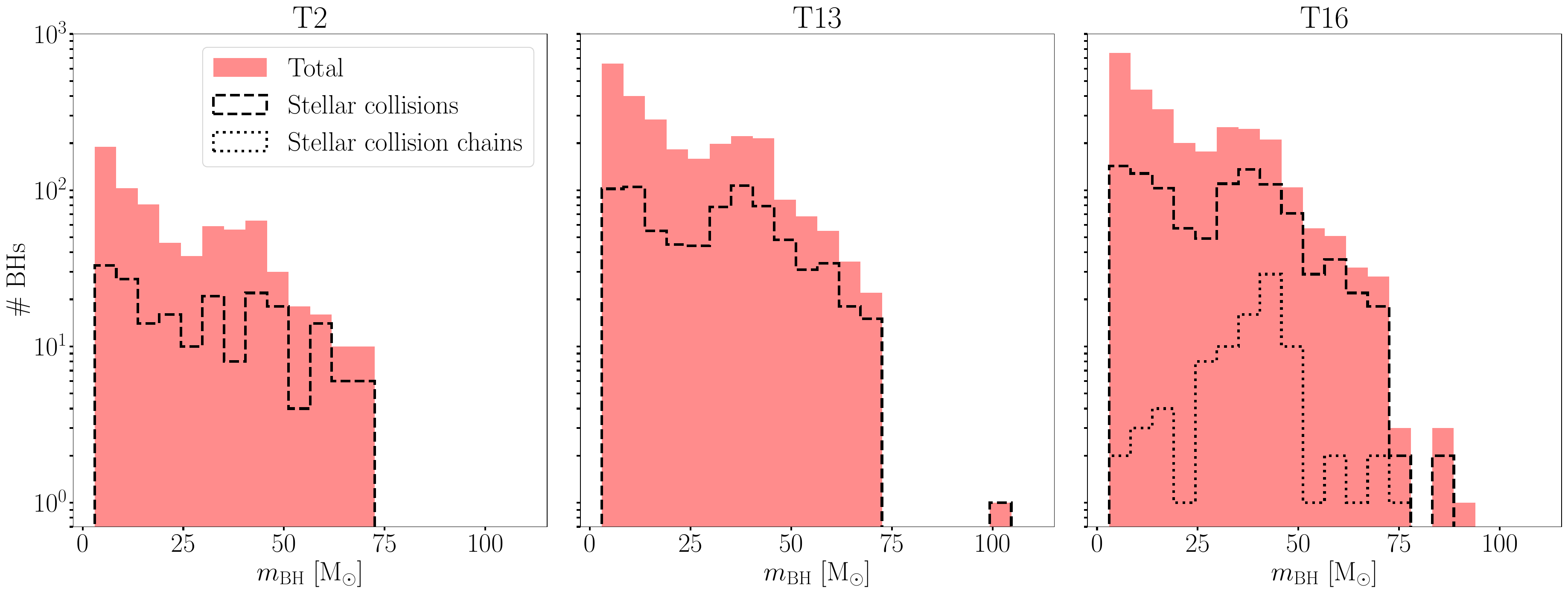}
    \caption{BH mass distribution for models T2, T13 and T16. We show in red the mass distribution of all BHs; the black dashed line shows the BHs formed through single stellar collisions; the black dotted line shows instead the BHs formed through repeated stellar collisions.}
     \label{fig:bh_hist}
\end{figure*}

To assess the impact of single and repeated stellar collisions on the BH mass spectrum, we focus on three models with different values of density and relaxation time (T2, T13, T16). Figure~\ref{fig:bh_hist} shows the corresponding BH mass distributions, highlighting the contributions from stellar collisions and stellar collision chains. As expected, the total number of BHs increases with the mass of the host cluster, ranging from 721 to 2892 BHs in the most massive model (T16). In parallel, the number of BHs originating from single stellar collisions also grows with cluster mass, accounting for $\sim 28-35\%$ of the total BH population \citep{dicarlo2019, mas2024}, reflecting the increasing number of massive stars and primordial binaries in more massive systems. A qualitatively different behavior is observed only in model T16, the densest cluster in our suite ($\rho_{\rm h}\sim10^5\,\msun\,\rm pc^{-3}$) with a short relaxation time ($t_{\rm rh}=300\,\rm Myr$). In this case, a non-negligible fraction ($3\%$) of BHs forms from the collapse of the final products of stellar collision chains.

The BH mass distributions are characterized by a pronounced primary peak at $m_{\rm BH}\sim5\,\msun$ and a secondary pile-up at $\sim35\,\msun$. The latter is primarily driven by the adopted prescriptions for pulsational PISNe \citep{woosley2017, spera2017}, which, at the low metallicity of our models, induce strong mass loss and effectively funnel BH remnants toward masses below $m_{\rm BH}\lesssim50\,\msun$. In addition, the large primordial binary fractions adopted in our simulations, combined with the preference for short initial orbital separations \citep{sana2012}, promote efficient envelope stripping and mass loss during binary evolution, further contributing to the accumulation of BHs in this mass range.

BHs formed from single stellar collisions broadly follow the overall mass distribution, with a comparatively smaller contribution to the low-mass peak. In contrast, BHs originating from stellar collision chains predominantly populate the $\sim35\,\msun$ peak. This behavior indicates that, even when stellar progenitors undergo substantial mass growth through repeated collisions, the combined effects of pulsational PISNe and binary interactions efficiently limit the final BH mass.
Finally, the number of BHs with $m_{\rm BH}\gtrsim60\,\msun$ is small, marking the lower edge of the PISN mass gap in our models. Nevertheless, we find that the population of BHs in the upper-mass gap grows with cluster mass. In model T13 a single IMBH forms; this case is discussed in detail in Sec.~\ref{sec:imbh}. In model T16, instead, the enhanced impact of stellar collisions and collision chains leads to the formation of 85 BHs with masses $60\leq m_{\rm BH}\lesssim90\,\msun$.

\subsection{Intermediate-mass black holes}\label{sec:imbh}
In the following, we consider two families of IMBHs ($m_{\rm BH}>100\,\msun$): those that arise from stellar collisions and those that originate from BBH mergers. These two classes are not mutually exclusive, as IMBHs formed through BBH mergers may have stellar progenitors that themselves experienced stellar collisions \citep{mas2023, mas2024b, rantala2025}.

\begin{figure*}[ht]
    \centering
    \includegraphics[width=0.7\textwidth]{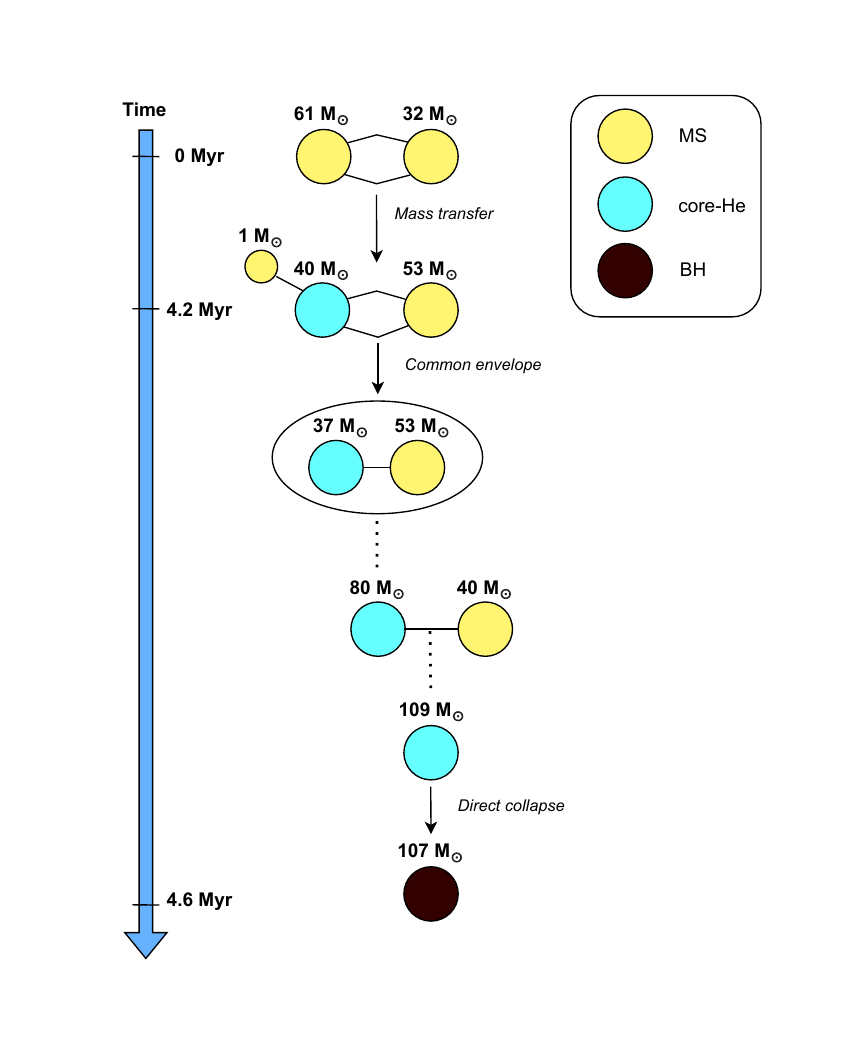}
    \caption{Formation pathway of the IMBH in model T9.}
     \label{fig:imbh_t10}
\end{figure*}

Among the 18 \textsc{titans} simulations that reached $t_{\rm sim}=20\,\rm Myr$, five produce an IMBH through a purely stellar channel. All these IMBHs have masses $\leq 107\,\msun$ and share the same formation pathway: a MS star merges with a cHeb companion, resulting in a star whose helium core remains below the threshold for pulsational PISN \citep{dicarlo2020, kremer2020b, mas2021}, and that will later collapse into an IMBH. In four models (T12, T13, T14, T18), this merger happens only once and between the components of a primordial binary. Model T9, however, exhibits a more complex evolution: the initial merger product (between a MS and a cHeb star) subsequently undergoes two additional collisions with MS stars (see Fig.~\ref{fig:imbh_t10}). 

By contrast, none of the VMSs formed in our clusters collapses into an IMBH. Instead, most VMSs either explode as PISNe (Sec.~\ref{sec:pisn}), leave behind stellar-mass BHs (Sec.~\ref{sec:bh}), or are stripped of mass through collisions with compact remnants (Appendix~\ref{app:fc_impact}). This means that, unless substantial material is accreted during BH-star collisions ($f_{\rm c}=0$) and within the model uncertainties, the seeding of IMBHs through the collapse of a VMS is not possible in star clusters with initial $\rho_{\rm h}\lesssim10^5\,\msun\,\rm pc^{-3}$ and $t_{\rm rh}>100\,\rm Myr$. 

We also assess the impact of stellar collisions on BBH mergers that produce IMBHs. For this analysis we consider the entire simulated duration of each model, and not just the first $20\,\rm Myr$. This choice introduces a bias toward simulations evolved for longer times, but it enables us to capture the full role of primordial binaries, stellar collisions and stellar collision chains across all clusters. Throughout the following discussion, we describe only IMBHs resulting from a single BBH merger, as relativistic kicks are not taken into account in \petar (see Sec.~\ref{sec:caveats}). We identify twelve IMBHs formed through BBH mergers, including four produced outside their parent cluster, all with masses $\lesssim 140\,\msun$. Among these, five have one progenitor produced by a primordial stellar merger, one has a progenitor that experienced repeated collisions, and one has both progenitors originating from primordial stellar mergers. As a consequence, we find that stellar mergers (T10, T12, T17) and mass transfer (T8) in primordial binaries can facilitate the formation of IMBHs through BBH mergers even before core collapse, when it is expected that high cluster densities promote the BBH coupling and eventual merger.

We note that most of the IMBHs formed in our simulations, whether originating from the stellar channel or from BBH mergers, lie within the PISN mass gap ($60 \lesssim m_{\rm BH} \lesssim 120\,\msun$). The high primordial binary fractions adopted in this study enable YMCs to produce IMBHs in, or just above, the PISN range through both stellar interactions and BBH mergers. However, the relatively low escape velocities of these clusters, combined with the absence of mechanisms to grow a seed above the PISN mass range, prevent the formation of substantially larger IMBHs. As a result, our simulations indicate that YMCs with properties consistent with those observed in the local Universe are unlikely to produce large IMBHs even in the presence of high primordial binary fractions.

\subsection{Caveats}\label{sec:caveats}

The initial conditions adopted in this work (see Table~\ref{table:ic_sims}) are broadly consistent with observations of YMCs in the local Universe. We also assume high primordial binary fractions \citep{moe2017} and employ a state-of-the-art treatment of single and binary stellar evolution. Nevertheless, our results rely on a number of assumptions that are worth discussing.

First, we assume that all clusters form through the monolithic collapse of a molecular cloud. In reality, star clusters may form from clouds with significant departures from spherical symmetry or through the hierarchical assembly of smaller sub-clusters \citep[e.g.,][]{sanchez2009, ballone2020,torniamenti2021}. Recent studies have shown that hierarchical assembly can significantly enhance the rate of repeated stellar collisions, favoring the formation of large ($\sim10^3\,\msun$) IMBH seeds \citep{rantala2024, rantala2025, rantala2026}. Our results should therefore be regarded as conservative with respect to scenarios in which cluster substructure persists during the early stages of evolution.

Second, the properties of YMCs may evolve with redshift. Recent observations with JWST have begun to reveal extremely compact and massive stellar systems in the early Universe (\citealp{adelaide2026} and references therein), characterized by high stellar densities and short relaxation times. Such environments would likely experience an enhanced efficiency of stellar collisions and collision chains, potentially allowing the formation of VMSs with helium cores above the PISN regime and increasing the probability of IMBH seed formation. For example, \cite{vergara2025} recently showed that a cluster with $\rho_{\rm h}\sim10^8\,\msun\,\rm pc^{-3}$ and $t_{\rm rh}\sim7\,\rm Myr$ forming in the early Universe can produce a VMS as massive as $50,000\,\msun$ collapsing in an IMBH.

Third, our conclusions depend on the adopted prescriptions for stellar and binary evolution. In particular, the stellar mass range associated with the PISN and pulsational PISN regimes is uncertain and model dependent. Alternative prescriptions based on \cite{spera2017} predict a narrower PISN mass range, which could allow IMBH seeds to form even under initial conditions similar to those explored here. Additional uncertainties arise from the treatment of stellar winds, mass transfer, and envelope stripping in massive binaries, all of which can affect the final masses of collision products and compact remnants. We also point out that in our work we have not considered collision-induced mass-loss \citep{gaburov2010, ramirez2025}.

We also assume that the fraction of stellar mass accreted by a BH during a collision is zero ($f_{\rm c}=0$). Adopting a higher accretion efficiency can significantly alter the outcomes of stellar-BH collisions and would lead to the formation of up to nine additional IMBHs in our simulations (see Appendix~\ref{app:fc_impact}).

Finally, \petar does not include relativistic recoil kicks following GW mergers. As a consequence, in Sec.~\ref{sec:imbh} we consider only first-generation BBH mergers as viable IMBH formation channels. Given the relatively low escape velocities of our clusters, we expect most successive-generation merger remnants to be ejected. A follow-up study will explore the long-term evolution of compact objects and their merger products in greater detail.

\section{Summary}\label{sec:summary}
We have investigated the impact and properties of repeated stellar collisions in YMCs over the first $20\,\rm Myr$ of evolution using the new set of direct $N$-body simulations \textsc{titans}. These models span a wide range of cluster masses and densities consistent with low-redshift YMCs and include a high, observation-based fraction of primordial binaries (Fig.~\ref{fig:init_cond}).

We find that the efficiency of repeated stellar collisions increases with increasing half-mass density $\rho_{\rm h}$ and decreasing half-mass relaxation time $t_{\rm rh}$ (Fig.~\ref{fig:eff_rhoh}). Clusters with $\rho_{\rm h}\gtrsim10^4\,\msun\,\rm pc^{-3}$ and dynamical friction times $t_{\rm rh}\lesssim300\,\rm Myr$ efficiently form stellar collision chains, although these typically involve only two collisions and are generally triggered by primordial binary mergers. Only a minority of chains in the densest clusters undergo three or more collisions (Fig.~\ref{fig:num_coll_chains}).

The evolutionary stage of the primary stars involved in repeated collisions depends strongly on the cluster relaxation time (Fig.~\ref{fig:run_k1}). In particular, the fraction of primaries on the MS decreases from $\sim77\%$ in the model with the shortest relaxation time (T1) to $\sim17\%$ in the model with the longest relaxation time (T17). The masses of primaries in collision chains vary accordingly, with typical values $50\lesssim m_1 \lesssim 130\,\msun$ in dense clusters with $t_{\rm rh}\leq700\,\rm Myr$ (Fig.~\ref{fig:run_m1}).

Very massive stars are predominantly formed via primordial binary mergers and typically remain below $300\,\msun$. Only in the densest clusters with the shortest relaxation times (T1, T9, T16) do repeated stellar collisions produce VMSs with masses up to $\sim400\,\msun$ (Fig.~\ref{fig:rho_max_vms}). As a consequence, the number of VMSs generally scales with cluster mass. Most of these stars either explode as PISNe or leave behind stellar-mass BHs with $m_{\rm BH}<50\,\msun$, and none represent viable IMBH seeds.

The number of PISNe in our clusters grows with cluster mass (Fig.~\ref{fig:num_pisn_mcl}), as the dominant formation channels involve massive stars and primordial binary mergers, whose frequency increases with cluster mass. Compared to an isolated population of single stars, which is expected to produce PISNe only for zero-age MS masses above $140\,\msun$, our clusters generate up to an order of magnitude more PISNe owing to the contribution of stellar collisions.

The BH mass distribution exhibits a primary peak at $\sim5\,\msun$ and a secondary peak at $\sim35\,\msun$ (Fig.~\ref{fig:bh_hist}). Stellar collision chains contribute mainly to the secondary peak, indicating that pulsational PISNe and binary interactions effectively regulate the final BH masses, even when stellar progenitors experience substantial mass growth through repeated collisions.

Finally, we find that IMBHs form through two main channels: (i) collisions, either single or repeated, between a cHeb star and a MS star (Fig.~\ref{fig:imbh_t10}), producing five IMBHs, and (ii) single BBH mergers, producing twelve IMBHs. In both cases, the resulting IMBHs remain relatively small ($m_{\rm BH}<140\,\msun$). Both formation pathways are strongly enhanced by the high primordial binary fraction, which promotes stellar collisions, mass transfer episodes, and the assembly of merging BBHs.

In summary, adopting initial conditions compatible with low-redshift YMCs and state-of-the-art $N$-body modeling, we find that repeated stellar collisions can be efficient when mass-segregation times are sufficiently short. However, collision chains typically involve only two events and are largely driven by primordial binaries. Within our conservative modeling assumptions ($f_{\rm c}=0$), VMSs do not act as viable IMBH seeds, and only collisions between stars at different evolutionary stages can produce small IMBH progenitors. While this work focuses on the early evolution of YMCs, our simulations are ongoing and have already reached $t_{\rm sim}>1\,\rm Gyr$ for the least dense clusters. A follow-up study will investigate the long-term evolution of compact objects and compact binary mergers in these systems.

\begin{acknowledgements}
MM acknowledges financial support from the European Research Council for the ERC Consolidator grant DEMOBLACK, under contract no. 770017. MM also acknowledges financial support from the German Excellence Strategy via the Heidelberg Cluster of Excellence (EXC 2181 - 390900948) STRUCTURES. 

MAS acknowledges funding from the European Union’s Horizon 2020 researchno  and innovation programme under the Marie Skłodowska-Curie grant agreement No.~101025436 (project GRACE-BH) and from the MERAC Foundation. MB and MAS acknowledge support from the Astrophysics Center for Multi-messenger Studies in Europe (ACME), funded under the European Union’s Horizon Europe Research and Innovation Program, Grant Agreement No. 101131928. 
SR acknowledges financial support from the Beatriu de Pinós postdoctoral fellowship program under the Ministry of Research and Universities of the Government of Catalonia (Grant Reference No. 2021 BP 00213).
The authors acknowledge support by the state of Baden-W\"urttemberg through bwHPC and the German Research Foundation (DFG) through grants INST 35/1597-1 FUGG and INST 35/1503-1 FUGG.

This research made use of \textsc{NumPy} \citep{harris2020}, \textsc{SciPy} \citep{scipy2020}, \textsc{Pandas} \citep{pandas2020}. For the plots we used \textsc{Matplotlib} \citep{hunter2007}. 
Our simulations made use of the $N$-body code \petar (\citealt{wang2020b} and \url{https://github.com/lwang-astro/PeTar}) and of the population-synthesis code \mobse (\citealt{mapelli2017,giacobbo2018b} and \url{https://gitlab.com/micmap/mobse_open}).
\end{acknowledgements}

\bibliographystyle{aa} 
\bibliography{bibliography.bib} 

\begin{appendix}

\section{Impact of the chosen accretion fraction $f_{\rm c}$}\label{app:fc_impact}
In the following we explore the impact of the assumed accretion fraction $f_{\rm c}$ in case of a collision or merger between a star and a BH.

\begin{figure*}[ht]
    \centering
    \includegraphics[width=0.9\textwidth]{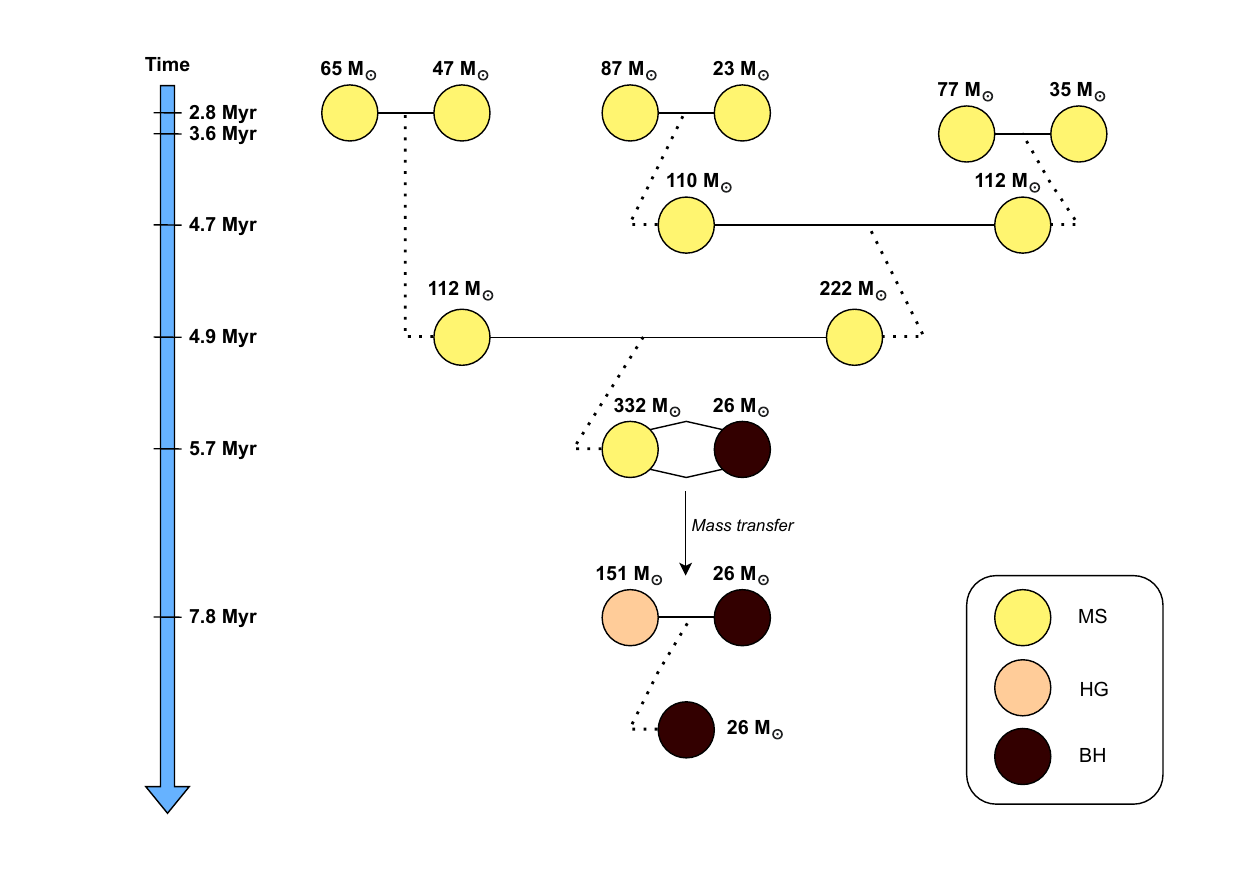}
    \caption{Stellar collision chain leading to a VMS-BH merger in model T1. In this case $f_{\rm c}=0$ and the final BH does not accrete mass from the VMS.}
     \label{fig:diag_vms_bh}
\end{figure*}

Table~\ref{table:vms_info} shows that nine VMSs in our set of simulations collide with a BH. Because of our assumption of $f_{\rm c}=0$, no mass is accreted by a BH in case of a collision with a star. A clear example of this scenario, where the VMS is the result of a stellar collision chain, is shown in Fig.~\ref{fig:diag_vms_bh}. In most cases, the VMSs form either via binary evolution, accreting mass from the companion which will later become a BH, or via a primordial binary merger. Once the companion is a BH, the star transfers its mass, which is lost, bringing away angular momentum and widening the orbit. The merger finally happens when the star expands enough.

If we assumed $f_{\rm c}=0.5$ (e.g. \citealp{mas2024}), we could obtain nine more IMBHs up to $m_{\rm BH}\sim215\,\msun$ in our simulations, deriving from the merger of binaries formed by a VMS and a BH. We also expect that the accretion of stellar mass from a VMS onto a stellar-mass BH might lead to a significant spin-up of the BH \citep{kiroglu2025, kiroglu2025b}. 

\begin{figure*}[ht]
    \centering
    \includegraphics[width=0.9\textwidth]{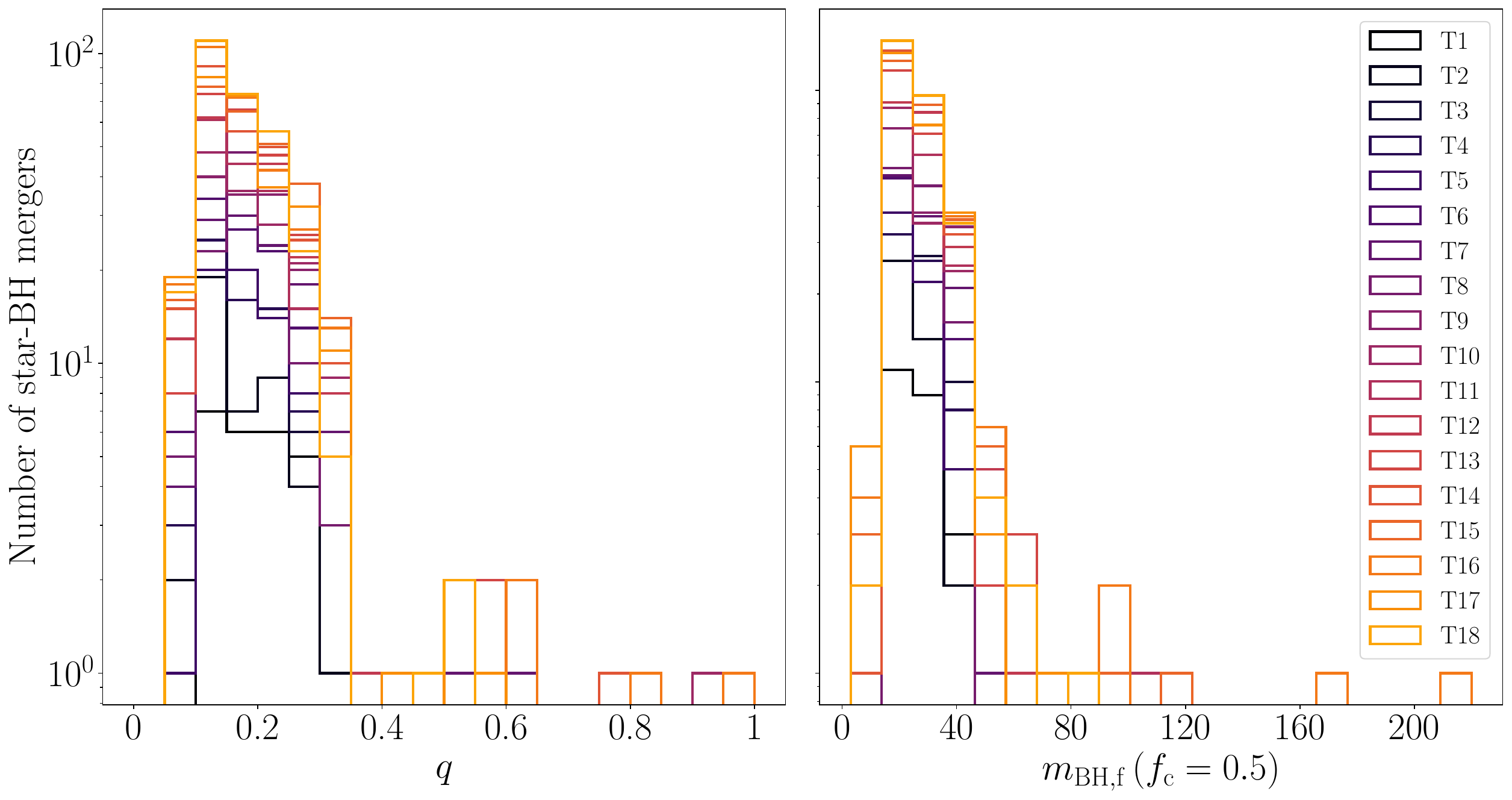}
    \caption{Mass-ratio $q$ distribution of merging systems with a stellar primary and a BH secondary (left) and final BH mass assuming $f_{\rm c}=0.5$ (right), for all the simulated models.}
     \label{fig:q_mf_star_bh}
\end{figure*}

To evaluate the impact of our prescription on the BH population in our models, we have also looked into all the collisions involving a primary star and a secondary BH. We show in Fig.~\ref{fig:q_mf_star_bh} the distribution of the mass ratio $q = m_{\rm BH}/m_{\rm *}$ for all the simulated models. We see that usually $q$ is smaller than $\sim0.4$ and that it shows a peak at $q\lesssim0.2$. Even though the amount of spin-up via stellar collisions is still largely unknown, we can expect that, in case of small mass ratios such as these, both the mass and spin of the BH would increase. Consequently, our adopted choice of $f_{\rm c}=0$ effectively represents a conservative scenario, as it minimizes the mass growth of BHs via stellar collisions in our models. Figure~\ref{fig:q_mf_star_bh} also shows the distribution of BH masses after collisions assuming $f_{\rm c}=0.5$. Most BHs in this case have final masses $\lesssim 50\,\msun$, with a peak around $m_{\rm BH,f}\sim 20\,\msun$. The only BHs reaching the IMBH regime originate from collisions in which the primary was a VMS; these cases are listed in Table~\ref{table:vms_info}.

\end{appendix}

\end{document}